\documentclass[showpacs,preprintnumbers,amsmath,amssymb]{revtex4}

\usepackage{amsmath}
\usepackage{mathrsfs}
\usepackage{}
\usepackage{graphicx}
\usepackage{subfigure}
\usepackage{amsfonts}
\usepackage{amssymb}%
\usepackage{color}

\newcommand{\be}{\begin{equation}}
\newcommand{\ee}{\end{equation}}
\newcommand{\bea}{\begin{eqnarray}}
\newcommand{\eea}{\end{eqnarray}}

\newcommand{\hf}{\frac{1}{2}}

\newcommand{\pa}{\partial}
\newcommand{\nn}{\nonumber\\}
\newcommand{\ie}{{\it i.e.}}
\newcommand{\etc}{{\it etc.}}
\newcommand{\eg}{{\it e.g.}}
\newcommand{\etl}{{\it et. al.}}

\providecommand{\Journal}[4] {#1 {\bf#2}, #4 (#3)}
\providecommand{\PLA}{Phys. Lett. A} %
\providecommand{\PRL}{Phys. Rev. Lett.} %

\providecommand{\PRA}{Phys. Rev. A}
\providecommand{\PRB}{Phys. Rev. B}
\providecommand{\PW}{Physics World}

\providecommand{\NT}{Nature}

\providecommand{\CEJP}{CEJP} %
\providecommand{\EL}{Europhys. Lett.} %
\providecommand{\NJP}{New J. Phys.} %
\providecommand{\RPP}{Rep. Prog. Phys.} %
\providecommand{\PRP}{Phys. Rep.} %
\providecommand{\RMP}{Rev. Mod. Phys.} %
\providecommand{\AJP}{Am. J. Phys.} %
\providecommand{\JAP}{J. Appl. Phys.} %

\begin{document}

\title{Revisiting 1-Dimensional Double-Barrier Tunneling in Quantum Mechanics}

\author{Zhi Xiao}
\email{blueseacat@126.com}
\author{Shi-sen Du}
\author{Chun-Xi Zhang}
\affiliation{Beihang University, Beijing 100191, China}

\begin{abstract}
This paper revisited quantum tunneling dynamics through a square double-barrier potential. We emphasized the similarity of tunneling dynamics through double-barrier and that of optical Fabry--P$\acute{e}$rot (FP) interferometer. Based on this similarity, we showed that the well-known resonant tunneling can also be interpreted as a result of matter multi-wave interference, analogous to that of FP interferometer. From this analogy, we also got an analytical finesse formula of double-barrier. Compared with that obtained numerically for a specific barrier configuration, we found that this formula works well for resonances at ``deep tunneling region".
Besides that, we also calculated standing wave spectrum inside the well of double barriers and phase time of double-barrier tunneling. The wave number spectrums of standing wave and phase time show another points of view on resonance. From semi-numerical calculations, we interpreted the peak of phase time at resonance as resonance life time, which coincides at least in order of magnitude with that obtained from uncertainty principle. Not to our surprise, phase time of double-barrier tunneling also saturates at long barrier length limit $l\rightarrow\infty$ as that of tunneling through a single barrier, and the limits are the same.
\end{abstract}
\maketitle

\section{Introduction}
Optical Fabry--P$\acute{e}$rot interferometer has a broad application in optics and has been already used in many optical apparatus, such as optical filter, spectrometer, single-mode resonant laser cavity, \etc. The most central feature of FP interferometer is its excellent filter effect. For example, a commercial FP interferometer can easily achieve a finesse of about 200 (though may not be fabricated with a two parallel plates configuration, the principle beneath are the same), and recently a finesse up to 130000 FP fiber interferometer was reported\cite{HFFFP}. The high finesse of FP cavity is accomplished through the multi-beam interference effect. On the other hand, interference is a crucial feature of quantum mechanics, so one may naturally wonder whether it is possible to construct a matter wave FP interferometer. The answer is yes, and there have already been several mesoscopic devices with operating principle similar to that of FP interferometer available in the laboratory, thanks to the advanced mesoscopic fabrication technology\cite{mesoscopic}.

The foundational design responsible for the realization of high finesse FP interferometer we think is the resonant cavity formed by two parallel plates.
To conceive a matter wave analogy, we think the well-known resonant tunneling is the most suited concept for this aim. Though we thought we had got an original idea of a matter wave FP interferometer, later we learned that Chamon \etl have already proposed such an idea much earlier, but with much more sophisticated concepts and interesting physics involved\cite{ChamonWen}. In their work, the tunneling particle is not non-interacting real particle, but low energy excitation--quasiparticle instead. And the double barriers we discuss below is replaced by two point contacts\cite{ChamonWen}. Their proposal is later called electronic or Hall FP interferometer and has received much more attention\cite{EHallFP} recently.
However, as our results is from a close analogy with optical FP interferometer, which require little calculation of Schr$\mathrm{\ddot{o}}$dinger equation provided we have known transmission and reflection coefficients of single barrier in advance, and with this optical analogy, we have got a finesse formula matching well with the numerical results for resonances at ``deep tunneling region", we think these results may be still valuable. In addition to calculate parameters relevant to double- barrier, the matter wave analogy of an optical FP interferometer, we have also obtained the tunneling time of this barrier with phase time approach. We find that this time saturates at the same limit as that of single barrier, and we interpret this quantity at the local maximum as the lifetime of resonance. Compared with numerical results obtained from uncertainty principle, we think this interpretation is reasonable and may find application in the calculation of decay rate of resonances, \eg, in cosmology\cite{RSTQFT}.

The paper is organized as follows. First, we give a brief introduction to the concept of quantum tunneling and FP interferometer in section \ref{1st}, emphasized the similarity involved in between. Then we present a calculation of tunneling through a double-barrier following the standard quantum mechanical approach, \ie, solving Schr$\mathrm{\ddot{o}}$dinger equation. Next we calculate the transmission and reflection coefficients by a close analogy with that of optical FP interferometer. All these are given in section \ref{2nd}. In section \ref{3rd}, we calculate the phase time of double-barrier potential. With our calculation, we show that numerically, one can associate phase time at resonant wave number with the lifetime of resonance reasonably. At last, we give a conclusion in section \ref{4th}.

%%%%%%%%%%%%%%%%%%%%%%%%%%%%%%%%%%%%%%%%%%%%%%%%%%%%%%%%%%%%%%%%%%%%%%%%%%%%%%%%%%%%%%%%%%%%%%%%%%%%%%%%%%%%%%%%%%%%%%%%%%%%%%%%%%%%%%%%%%%%%%%%%%%%%%%
%It needs some extension to include more recent development on the subject and also needs some clear clue on how to organize this paper, including its
%novel points, the purpose of this paper, its conclusion.
%%%%%%%%%%%%%%%%%%%%%%%%%%%%%%%%%%%%%%%%%%%%%%%%%%%%%%%%%%%%%%%%%%%%%%%%%%%%%%%%%%%%%%%%%%%%%%%%%%%%%%%%%%%%%%%%%%%%%%%%%%%%%%%%%%%%%%%%%%%%%%%%%%%%%%%
\section{Quantum Tunneling}\label{1st}
\begin{figure}
 \centering\subfigure[~Square potential barrier]{\label{SSP}\includegraphics[width=60mm]{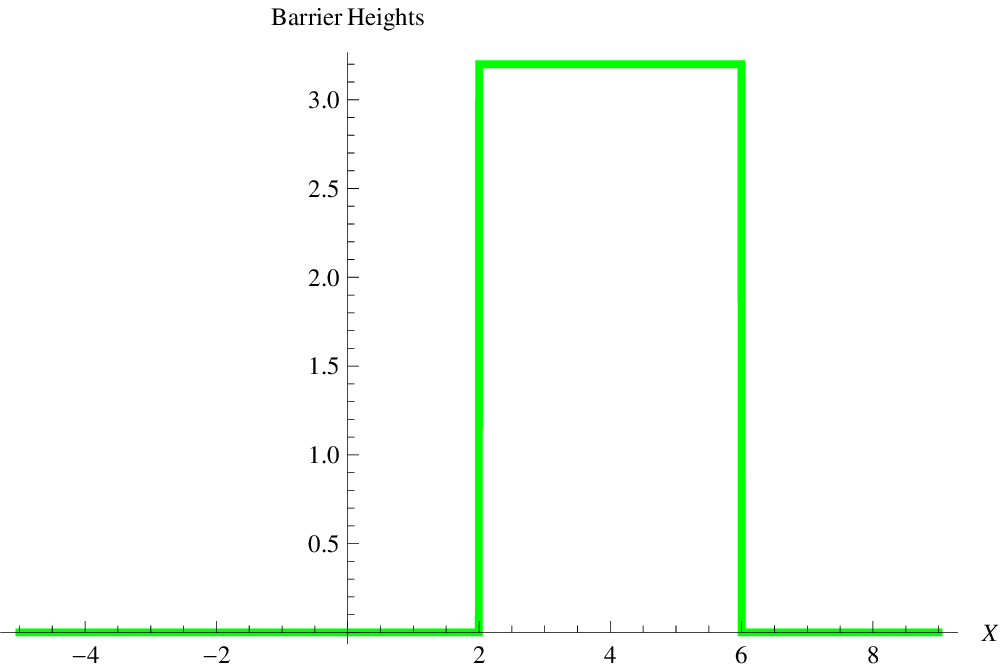}}
  \hspace{0.1in}
  \subfigure[~Double Square potential barriers]{\label{DSP}\includegraphics[width=60mm]{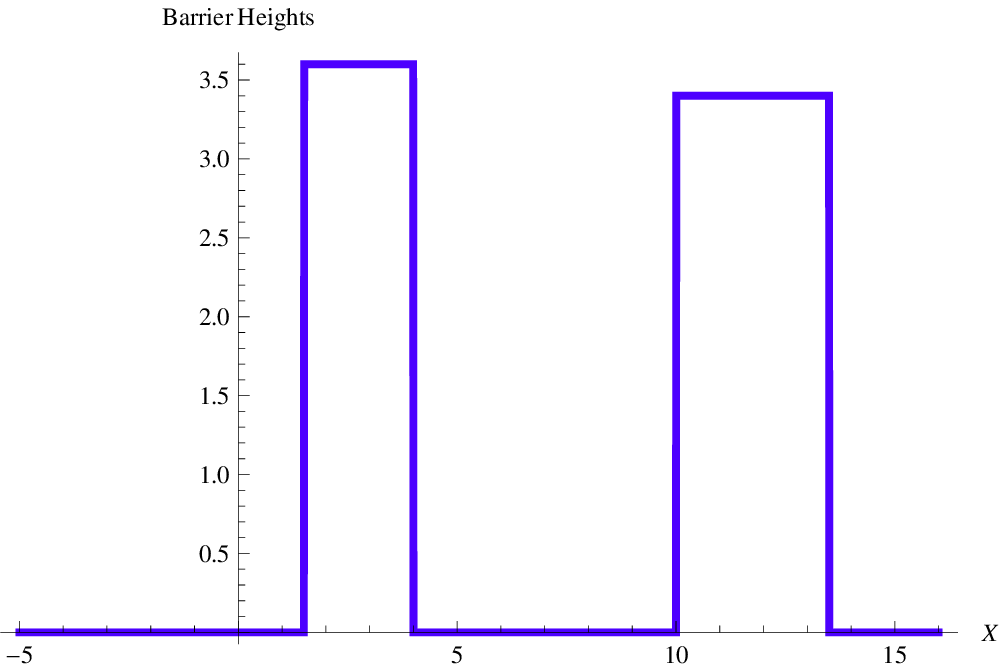}}
  \caption{(a)Square potential barrier with heights $A=3.2$eV, length $l=4$nm; (b). Double square potential barriers with heights $A_1=3.6$eV, length $l_1=2.5$nm and $A_2=3.4$eV, length $l_2=3.5$nm, two barriers are separated by a distance $d=6$nm.}\label{SquareB}
\end{figure}

In this section, we give an illustrative example of quantum tunneling, which shows some similarity with that of optical FP interferometer and also inspired us to find a
much resemblant matter wave counterpart. Consider the problem of a beam of particles (\eg, electrons or neutrons) with energy $\hbar\omega(k)=\frac{(\hbar{k})^2}{2m}$ scattered by the simplest square potential barrier of the form $V(x)=A[\Theta(x-a)\Theta(b-x)]$ (see Fig.\ref{SSP}),

the solution of the corresponding Schr$\mathrm{\ddot{o}}$dinger equation is
\bea\label{steepBarr}&&
\Psi(x,t)=\phi(x)e^{-i\omega(k)t}\\&&
\mathrm{with}~~~
\phi(x)=\left\{\begin{array}{c}
          \sqrt{I_0}(e^{ikx}+\mathcal{R}e^{-ikx}),\quad   ~~~~~~~~~x<a; \\
          \sqrt{I_0}(ce^{\beta{x}}+de^{-\beta{x}}), \quad  ~~~a<x<b; \\
          \sqrt{I_0}\mathcal{T}e^{ikx},\quad   ~~~~~~~~~~~~~~~~~~~~~~~x>b.
        \end{array}\right.
\eea
Where $\beta^2=\frac{2m}{\hbar^2}(A-\hbar\omega(k))$, $I_0$ is the incident flux, and the transmission and reflective coefficients are given by
\bea\label{TTCoeff}&&
\mathcal{T}=\frac{{e^{ik(a-b)}}}{\frac{i}{2}(\frac{\beta}{k}-\frac{k}{\beta})\sinh[\beta(b-a)]+\cosh[\beta(b-a)]},\\
&&\label{RTCoeff}
\mathcal{R}=\frac{-\frac{i}{2}(\frac{\beta}{k}+\frac{k}{\beta})\sinh[\beta(b-a)]e^{2ika}}{\frac{i}{2}(\frac{\beta}{k}-\frac{k}{\beta})\sinh[\beta(b-a)]+\cosh[\beta(b-a)]}.
\eea

It is interesting to note that if we make a sign change, \ie, $A\rightarrow-A$ and $\beta\rightarrow{i\beta}$, then potential barrier becomes potential well. Using the well-known relations
\bea
\sinh(i\beta{x})=i\sin{\beta{x}}, \quad ~~\cosh(i\beta{x})=\cos{\beta{x}},
\eea

the transmission rate (the square modulus of $\mathcal{T}$) now becomes
\bea\label{Trate}
T_{sw}=|\frac{2k\beta{e^{-ikL}}}{2k\beta\cos[\beta{L}]-i(\beta^2+k^2)\sin[\beta{L}]}|^2=\frac{1}{1+\frac{1}{4}(\frac{k}{\beta}-\frac{\beta}{k})^2\sin^2[\beta{L}]}
, \quad\text{with}~~L=b-a.
\eea
We note that the transmission rate is a periodic function of ${\beta}L$. Under certain circumstances, the transmission rate is equal to 1, in other words, the incident beam of particles completely transmit through the well. This is the well-known resonant transmission phenomena\cite{ZJY}.

This phenomena bears some similarity with a photon beam transmitted through a Fabry--P$\acute{e}$rot (FP) interferometer, though in the latter case the transmission rate is a periodic function of photon energy $E_\lambda$ or wave number $k$, \ie,
\bea\label{FPformu}
T_{FP}=\frac{1}{1+(\frac{2\mathfrak{F}}{\pi}\sin[\frac{2\pi{d_0}}{\lambda}])^2}=[1+(\frac{2\mathfrak{F}}{\pi}\sin[\frac{E_\lambda{d_0}}{c\hbar}])^2]^{-1},
\eea
with
\bea\label{FPfine}
 \mathfrak{F}=\pi\sqrt{\mathscr{R}}/(1-\mathscr{R}), \quad E_\lambda=\hbar\omega=hc/\lambda.
\eea
$\mathfrak{F}$ defined in (\ref{FPfine}) is the finesse of FP interferometer, and only depends on reflectivity $\mathscr{R}$ in this case.
Note that if $1-\mathscr{R}\ll1$, the transmission rate is very small except that incoming photon is at resonant energy $E_\lambda=\frac{c{\hbar}m\pi}{d_0}$  (where $m\in\mathcal{Z}$); while for a quantum well, it was supposed to be 1 in classical logic but is shown to be less than 1 in quantum mechanics in general except at resonances. Of course, as a quantum analogy of optical FP interferometer, quantum well indeed captures the essential concept (resonant transmission) and many interesting works have devoted to this problem\cite{CQFP}, however, we do not think this analogy is the most suited one. We find maybe tunneling through a double-barrier is a more suited analogy for this study.
In the next section, we will calculate the tunneling rate through a double-barrier configuration, in which the physic picture bears much resemblance than that of a quantum well to the optical FP interferometer's. In addition to the standard quantum mechanical calculation, we give another derivation of the transmission and reflection coefficients based on the close analogy between optical FP interferometer and a double-barrier potential.

\section{Double-Barrier Potential}\label{2nd}
\subsection{General Result}\label{general}
The solution of a beam of particles scattered by a double-barrier potential
\bea\label{DPB}&&
U_{db}(x)=A_1[\Theta(x-a_1)\Theta(b_1-x)]+A_2[\Theta(x-a_2)\Theta(b_2-x)],\quad~\text{with}~~b_i-a_i=l_i,~i=1,2;\quad a_2-b_1=d.
\eea
can be obtained by solving the corresponding Schr$\mathrm{\ddot{o}}$dinger equation
\bea\label{Scheom}
i\hbar\frac{\pa\Psi(x,t)}{\pa{t}}=[-\frac{\hbar^2}{2m}\frac{\pa^2}{\pa{x^2}}+U_{db}(x)]\Psi(x,t),
\eea
where $d$ is the distance between double potential barriers and $l_1,~l_2$ and $A_1,~A_2$ are the widths and heights of the corresponding barriers respectively.
The double-barrier configuration is shown in Fig.\ref{DSP}. The two-barrier is reminiscent of the parallel two plates in FP interferometer, later we will use this resemblance to derive the transmission and reflection coefficient in another way.

With the given ansatz
\bea\label{steepBarr}&&
\Psi(x,t)=\phi(x)e^{-i\omega(k)t}\\&&
\mathrm{with}~~~\label{ACoeff}
\phi(x)=\left\{\begin{array}{c}
          e^{ikx}+\mathcal{R}_{db}e^{-ikx},\quad   ~~~~~~~~~~~~~~~~~~~~~~~x<a_1; \\
          c_1e^{\beta_1{x}}+d_1e^{-\beta_1{x}}, \quad  ~~~~~~~~~~~~~~~a_1<x<b_1; \\
          \alpha{e}^{ikx}+\gamma{e}^{-ikx},\quad   ~~~~~~~~~~~~~~~~b_1<x<a_2; \\
          c_2e^{\beta_2{x}}+d_2e^{-\beta_2{x}}, \quad  ~~~~~~~~~~~~~~~a_2<x<b_2; \\
          \mathcal{T}_{db}e^{ikx}, \quad  ~~~~~~~~~~~~~~~~~~~~~~~~~~~~~~~~~~x>b_2, \\
        \end{array}\right.
\eea
(where we have normalized incident flux $I_0$ to 1) and the continuity condition of wave-function
\bea\label{Continuity1}&&
\phi(a_i-0)=\phi(a_i+0),\quad ~~\phi(b_i-0)=\phi(b_i+0),\\&&\label{Continuity2}
\frac{d\phi(x)}{dx}|_{x=a_i-0}=\frac{d\phi(x)}{dx}|_{x=a_i+0},\quad ~~\frac{d\phi(x)}{dx}|_{x=a_i-0}=\frac{d\phi(x)}{dx}|_{x=a_i+0},
~~~~~~ (i=1,2)
\eea
we can obtain the transmission coefficient
\bea\label{Tdcoeff}&&
\mathcal{T}_{db}=\frac{e^{-ik(d+l_1+l_2)}}{\mathcal{F}[k;\beta_1,\beta_2;d;l_1,l_2]},\\&&\label{Fcoef}
\mathcal{F}[k;\beta_1,\beta_2;d;l_1,l_2]=e^{ikd}[M_1\sinh(\beta_1{l_1})][M_2\sinh(\beta_2{l_2})]+
e^{-ikd}[\cosh(\beta_1{l_1})+iK_1\sinh(\beta_1{l_1})]\nn&&
~~~~~~~~~~~~[\cosh(\beta_2{l_2})+iK_2\sinh(\beta_2{l_2})],
\eea
where $M_i,~K_i$ in (\ref{Fcoef}) are defined as
\bea\label{defMK}
M_i\equiv\hf(\frac{\beta_i}{k}+\frac{k}{\beta_i}), \quad K_i\equiv\hf(\frac{\beta_i}{k}-\frac{k}{\beta_i});  ~~~(i=1,2).
\eea
The other coefficients are given explicitly in Appendix A.\\

\begin{figure}
        \begin{center}
         \scalebox{1.0}{\includegraphics[width=70mm]{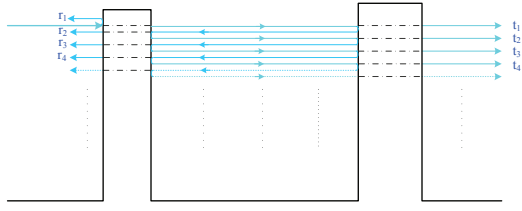}}
          \caption{\scriptsize A matter multi-wave interference for a beam of particles tunneling through a double-barrier potential.} \label{MultiWave}
        \end{center}
\end{figure}

We have already observed the formal analogy between two barriers with the two parallel plates in FP interferometer. Next we need to know the reflection and transmission rate and the corresponding phase changes for a particle tunneling through or reflected by each barrier respectively. These have already been obtained in section \ref{1st}, here we decompose the amplitudes
\bea\label{Tsrate}&&
T_i=|\mathcal{T}_i|=\frac{1}{[1+M_i^2\sinh^2(\beta_i{l_i})]^{1/2}}, \\&&\label{Rsrate}
R_i=|\mathcal{T}_i|=\frac{M_i\sinh(\beta_i{l_i})}{[1+M_i^2\sinh^2(\beta_i{l_i})]^{1/2}}; ~~~(i=1,2)
\eea
and phases
\bea\label{Tspha}&&
\phi_{ti}=-kl_i-\phi_i, \\&&\label{Rspha}
\phi_{ri}=-\pi/2-\phi_i, \quad \phi_i=\arctan[K_i\tanh(\beta_i{l_i})]; ~~~(i=1,2)
\eea
of the corresponding formulas (\ref{TTCoeff},\ref{RTCoeff}) explicitly.
Next we will use this optical FP analogy, or more precisely multi-wave interference (shown in Fig.\ref{MultiWave}) to regain the transmission and reflection coefficients of double barriers.

For a beam of particle scattering on a double-barrier potential, it will be partially reflected and partially transmitted through the first barrier. The reflected component is denoted as $r_1$; the transmitted component propagates forward through a distance $d$ and meets the second barrier. Then it will partially transmitted and partially reflected again, the transmitted component is denoted as $t_1$, while the reflected component propagates backward through another distance $d$ and then meets the first barrier. Then it will partially transmitted with transmitted component denoted as $r_2$, and the reflected component will repeat the story again and again, as shown in Fig.\ref{MultiWave}. The final amplitude of transmitted and reflected waves are the summation of these transmitted and reflected partial waves, \ie
\bea\label{Tdcoef}&&
\mathcal{T}_{db}=\sum_{i=1}^{+\infty}t_i=T_1T_2\exp[i(\phi_{t1}+\phi_{t2})+ik(d+l_1+l_2)]+T_1R_2R_1T_2\exp[i(\phi_{t1}+\phi_{r1}+\phi_{r2}+\phi_{t2})+ik(3d+l_1+l_2)]\nn&&
~~~~~~+T_1(R_2R_1)^2T_2\exp\{i[\phi_{t1}+2(\phi_{r1}+\phi_{r2})+\phi_{t2}]+ik(5d+l_1+l_2)\}+\ldots\nn&&
~~~~=T_1T_2\exp[i(\phi_{t1}+\phi_{t2})+ik(d+l_1+l_2)]\frac{1}{1-R_1R_2\exp[i(\phi_{r1}+\phi_{r2}+2kd)]}\nn&&
~~~~=\frac{-\exp\{i[\phi_{t1}+\phi_{t2}-(\phi_{r1}+\phi_{r2})+k(l_1+l_2)]\}}{\frac{R_1R_2}{T_1T_2}\exp(ikd)-\frac{1}{T_1T_2}\exp[-i(\phi_{r1}+\phi_{r2}+kd)]}
\eea
From (\ref{Tsrate},\ref{Rsrate}) and (\ref{Tspha},\ref{Rspha}), we can obtain
{\footnotesize
\bea\label{ratio1}
\frac{R_1R_2}{T_1T_2}=M_1\sinh(\beta_1{l_1})M_2\sinh(\beta_1{l_2}), \quad -\frac{\exp[-i(\phi_{r1}+\phi_{r2})]}{T_1T_2}=\{[1+M_1^2\sinh^2(\beta_1{l_1})][1+M_2^2\sinh^2(\beta_2{l_2})]\}^{1/2}\exp[i(\phi_1+\phi_2)]
\eea
}
and the phase factor
\bea\label{phast}
\phi_{t1}+\phi_{t2}-(\phi_{r1}+\phi_{r2})+k(l_1+l_2)=\pi.
\eea
Substituting (\ref{ratio1},\ref{phast}) into (\ref{Tdcoef}) and utilizing the identity
\bea\label{docomComp}
\frac{1}{T_i}\exp(i\phi_i)=[1+M_i^2\sinh^2(\beta_i{l_i})]^{1/2}\exp\{i\arctan[K_i\tanh(\beta_i{l_i})]\}=\cosh[\beta_i{l_i}]+iK_i\sinh[\beta_i{l_i}], ~~~(i=1,2)
\eea
we can obtain the transmission coefficient
\bea\label{redTdco}&&
\mathcal{T}_{db}=\frac{1}{\mathcal{F}[k;\beta_1,\beta_2;d;l_1,l_2]}.
\eea
Compared with (\ref{Tdcoeff}), there missed a phase factor $e^{-ik(d+l_1+l_2)}$. This phase factor can be inserted back from the comparison with single barrier case (\ref{TTCoeff}) in the limit $d\rightarrow0$ (which picks up a phase factor $e^{-ik(l_1+l_2)}$) and that with right moving plane wave $e^{ikx}$ in the thin barrier limit $l_1=l_2\rightarrow0$.

Similarly, the reflective coefficient can be obtained with the same procedure utilizing multi-wave interference, and the result coincides exactly with (\ref{Rcoeffd}) calculated from Schr$\mathrm{\ddot{o}}$dinger equation. For details, see Appendix B.\\

From the derivation of transmission coefficient (\ref{redTdco}), we can immediately get the resonant tunneling\cite{ZJY}\cite{CQFP}\cite{ResonantTun} condition by observing that
\bea\label{ResonCD}
|\frac{1}{1-R_1R_2\exp[i(\phi_{r1}+\phi_{r2}+2kd)]}|^2=\frac{(1-R_1R_2)^{-2}}{1+\frac{4R_1R_2}{(1-R_1R_2)^2}\sin^2[kd+\hf(\phi_{r1}+\phi_{r2})]},
\eea
hence the transmission rate gets the maximum value
\bea\label{maximum}
T_{max}=(\frac{T_1T_2}{1-R_1R_2})^2
\eea
only when
\bea\label{RESC}
kd+\hf(\phi_{r1}+\phi_{r2})=m\pi, ~~m\in\mathcal{Z}.
\eea
Note this condition, \ie, $\Phi(k)\equiv{kd-\hf(\phi_1+\phi_2+\pi)}=m\pi,~~m\in\mathcal{Z}$ is obtained by a direct analogy with the expression of FP interferometer, (\ref{FPformu}).
From the same formal analogy, we can also define the finesse of a double-barrier interferometer as
\bea\label{fineTHEO}
\mathfrak{F}_{db}=\frac{\pi\sqrt{R_1R_2}}{1-R_1R_2}=\frac{\pi[M_1\sinh(\beta_1{l_1})M_2\sinh(\beta_2{l_2})]^\hf\{[1+M_1^2\sinh^2(\beta_1{l_1})][1+M_2^2\sinh^2(\beta_2{l_2})]\}^{1/4}}
{\{[1+M_1^2\sinh^2(\beta_1{l_1})][1+M_2^2\sinh^2(\beta_2{l_2})]\}^{1/2}-M_1\sinh(\beta_1{l_1})M_2\sinh(\beta_2{l_2})}.
\eea
This formula works well especially for resonance at ``deep tunneling region", \ie, the peak of resonance in wave number spectrum is very sharp. This will be confirmed numerically in below. From (\ref{maximum}) we know in general $T_{max}<1$, except $T_1=T_2$. The simplest case where this condition is fulfilled is the symmetric double-barrier. We will utilize this special case (\ie, $l_1=l_2=l$ and $A_1=A_2=A$) to give a numerical calculation of finesse and resonance condition in the following. The numerical calculation in turn serves as a cross check of the usefulness of our analogy.

\subsection{Symmetric Double-Barrier}\label{special}
In this case, the transmission coefficient is
\bea\label{TranscDSPS}&&
\mathcal{T}_{dbs}=\frac{e^{-ik(d+2l)}}{\mathcal{F}[k,\beta,l,d]},
\eea
with
\bea&&\label{InFDSPS}
\mathcal{F}[k,\beta,l,d]=[M\sinh(\beta{l})]^2e^{ikd}+[iK\sinh(\beta{l})+\cosh(\beta{l})]^2e^{-ikd},
\eea
where
\bea\label{defMKf}
M\equiv\hf(\frac{\beta}{k}+\frac{k}{\beta}), \quad K\equiv\hf(\frac{\beta}{k}-\frac{k}{\beta}).
\eea

The transmission rate is given by
\bea&&\label{TdrateS}
T_{dbs}=|\mathcal{T}_{dbs}|^2=\frac{1}{|\mathcal{F}[k,\beta,l,d]|^2},
\eea
with

{\footnotesize
\bea\label{TdrateS2^-1}&&
|\mathcal{F}[k,\beta,l,d]|^2%=1+M^2[\cosh(2\beta{l})-1]\{1+\frac{M^2}{2}[\cosh(2\beta{l})-1]\}\nn&&~~~~+M^2[\cosh(2\beta{l})-1]\{[\cosh(\beta{l})^2-K^2\sinh(\beta{l})^2]
%\cos(2kd)+2K\sinh(\beta{l})\cosh(\beta{l})\sin(2kd)\}\nn&&~~~
=1+M^2[\cosh(2\beta{l})-1]\left[1+\frac{M^2}{2}[\cosh(2\beta{l})-1]+K\sinh(2\beta{l})\sin(2kd)+\{\cosh(2\beta{l})-\frac{M^2}{2}[\cosh(2\beta{l})-1]\}\cos(2kd)\right]\nn&&
~~~=1+M^2[\cosh(2\beta{l})-1]\left\{1+\frac{M^2}{2}[\cosh(2\beta{l})-1]\right\}[1+\sin(2kd+\delta)],
\eea
}
where in the last step of (\ref{TdrateS2^-1}) we have defined a $k$-dependent angle
{\footnotesize
\bea\label{redefine}&&
\tan\delta\equiv\frac{\cosh(2\beta{l})-\frac{M^2}{2}[\cosh(2\beta{l})-1]}{K\sinh(2\beta{l})}
\eea
}
and utilized an identity
{\footnotesize
\bea\label{iden}&&
[K\sinh(2\beta{l})]^2+\{\cosh(2\beta{l})-\frac{M^2}{2}[\cosh(2\beta{l})-1]\}^2\equiv\left[1+\frac{M^2}{2}[\cosh(2\beta{l})-1]\right]^2.
\eea
}

From (\ref{TdrateS},\ref{TdrateS2^-1}), it is easy to find that when $2kd+\delta=2N\pi+\frac{3\pi}{2}$,
$T_{dbs}=1$, \ie, incoming particles completely transmitted. While for the case
$2kd+\delta=2N\pi+\frac{\pi}{2}$, the transmission rate is locally minimal, and corresponds to the minimal tunneling
\bea\label{minimal}
T_{min}=(\frac{T_1T_2}{1+R_1R_2})^2
\eea
in section \ref{general}.

The above two local extremal conditions can be summarized as
\bea\label{localextrem}
\tan[\delta(k)]=\cot(2kd).
\eea
%One can verify that this condition is a special case (symmetric parameters) of resonant condition (\ref{RESC}) and the local minimum condition
%\bea\label{minimum}
%kd-\hf(\phi_1+\phi_2)=m\pi, ~~m\in\mathcal{Z}
%\eea
%shown in section \ref{general}, where $(\phi_1+\phi_2+\delta)\text{mod}\pi=\pi/2$ if $\phi_1=\phi_2$.

To determine at which wave-vector $k$ there is a resonance, we have to solve simultaneous equations (\ref{localextrem}) and
\bea\label{resonance}
\sin[2kd]=-\frac{K\sinh(2\beta{l})}{1+\frac{M^2}{2}[\cosh(2\beta{l})-1]}.
\eea
However, these equations (including the simultaneous equations to determine local minimal of transmission) are transcendental equations, so we can only solve them numerically.

\begin{figure}
        \begin{center}
         \scalebox{1.0}{\includegraphics[width=50mm]{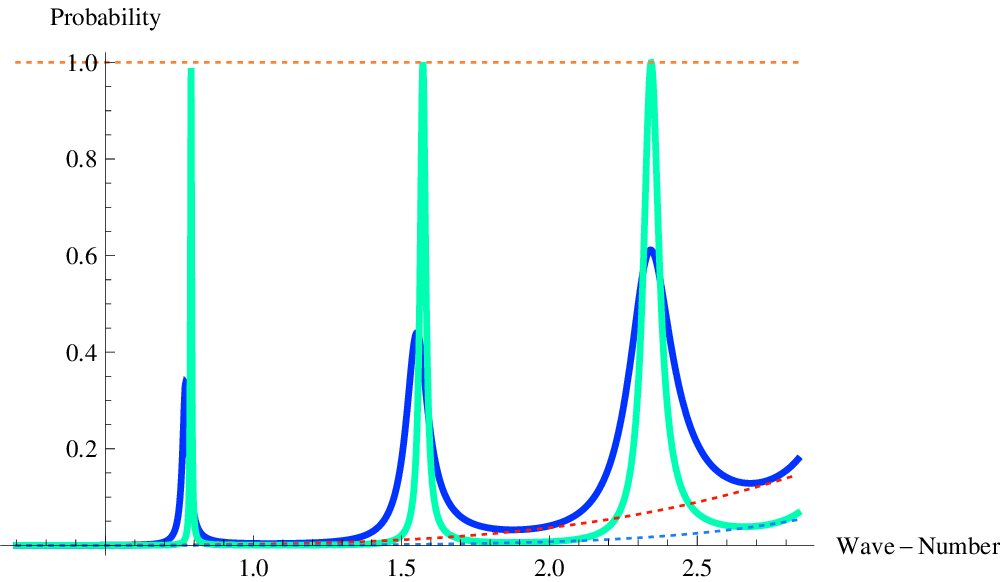}}
         \scalebox{1.0}{\includegraphics[width=50mm]{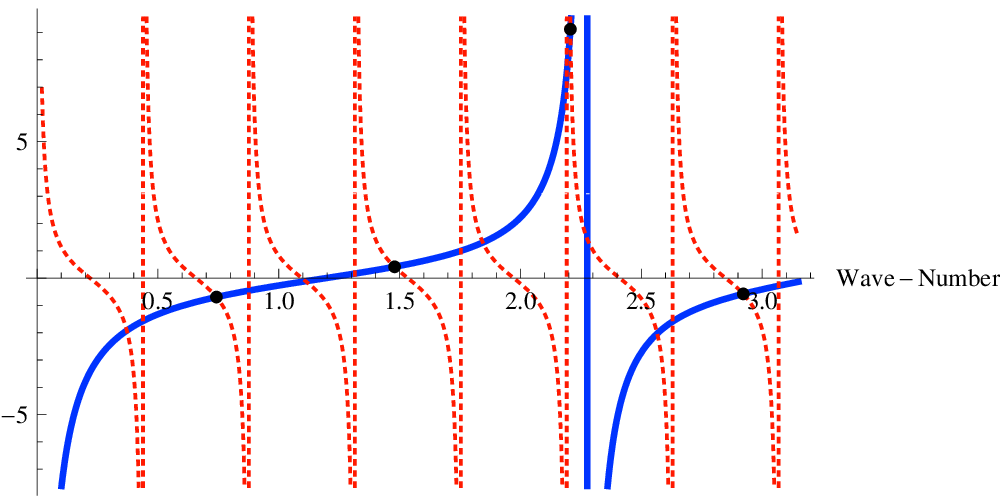}}
         \scalebox{1.0}{\includegraphics[width=50mm]{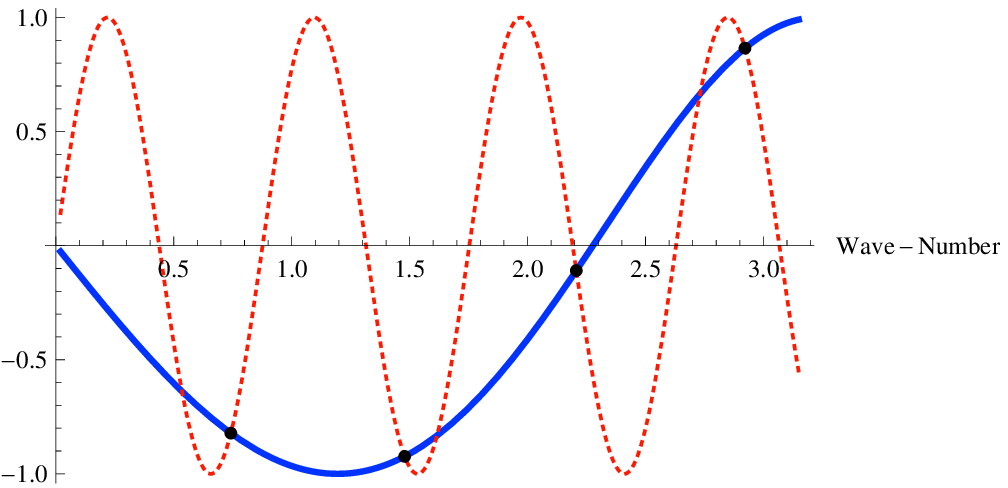}}
        \caption{\scriptsize left figure shows transmission rate with respect to the incoming particle wave-number $k$.
                    The green curve is plotted with symmetric double barriers: heights $A=10.36$eV, length $l=1.2{\AA}$ and distance $d=6.5{\AA}$. The blue one are two identical curves plotted with pair of parameters ($A_1=10.36$eV, $l_1=1.2{\AA}$) and ($A_2=9.6$eV, $l_2=0.6{\AA}$) interchanged, the distance between two barriers is also $d=6.5{\AA}$. The middle and right figures with cross points corresponding to the roots of two transcendental equations of (\ref{localextrem}) and (\ref{resonance}) respectively, and those indicated by dark disks are roots of the simultaneous equations. Note horizontal axis of wave number is plotted with unit $k_e=\sqrt{2m_eE_s}/\hbar$, \ie, the wave number of an electron with kinematic energy $E_s=1$eV.} \label{Asymm}
        \end{center}
\end{figure}

The results are illustrated in Fig.\ref{Asymm}. In the left subgraph of Fig.\ref{Asymm}, we have plotted the curves of transmission rate (\ref{Tdcoeff}) vs incoming wave-number $k$. The blue curve shows the transmission spectrum of general asymmetric barriers (with parameters $\beta_1\neq\beta_2$ and $l_1\neq{l_2}$). Note that this is actually a layered curve of two identical ones, with pairs of parameters $A_1\leftrightarrow{A_2}$, $l_1\leftrightarrow{l_2}$ interchanged simultaneously. This permutation symmetry is manifested in the transmission formula (\ref{Fcoef}), and inherits from the left-right symmetry of 1-Dim Schrodinger equation (\ref{Scheom}), \ie, particles incoming from left must have the same transmission rate with those coming from right. As we already stressed in the end of section \ref{general}, the maximum transmission rate in the asymmetric-barrier case is less than 1 in general. While for the symmetric case, the green curve, its peaks approach unit and provides a much ideal example of optical FP analogy.
\begin{figure}
 \centering
  \subfigure[~Transmission Spectrum for Optical FP Interferometer]{\label{FPcurve}\includegraphics[width=70mm]{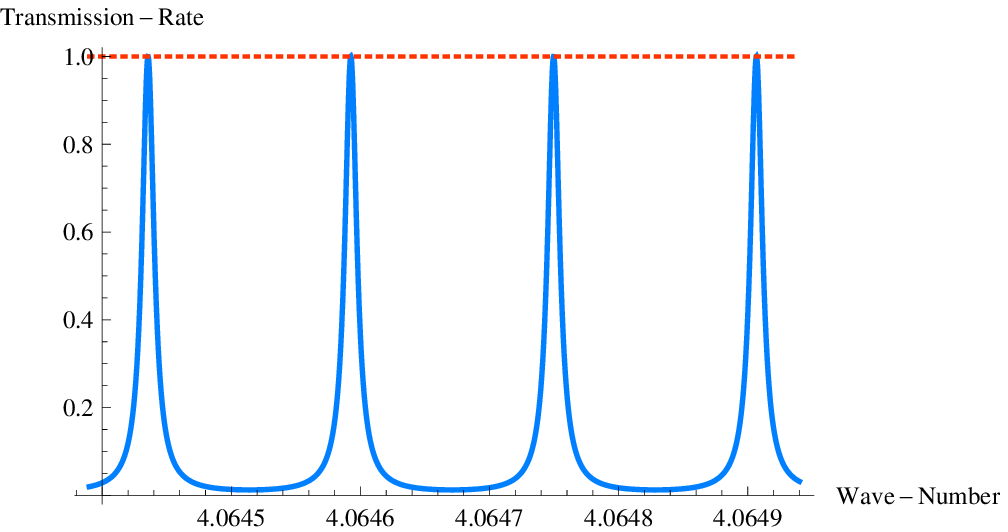}}
  \hspace{1in}
  \subfigure[~Transmission Spectrum for Double Barrier]{\label{finesse}\includegraphics[width=65mm]{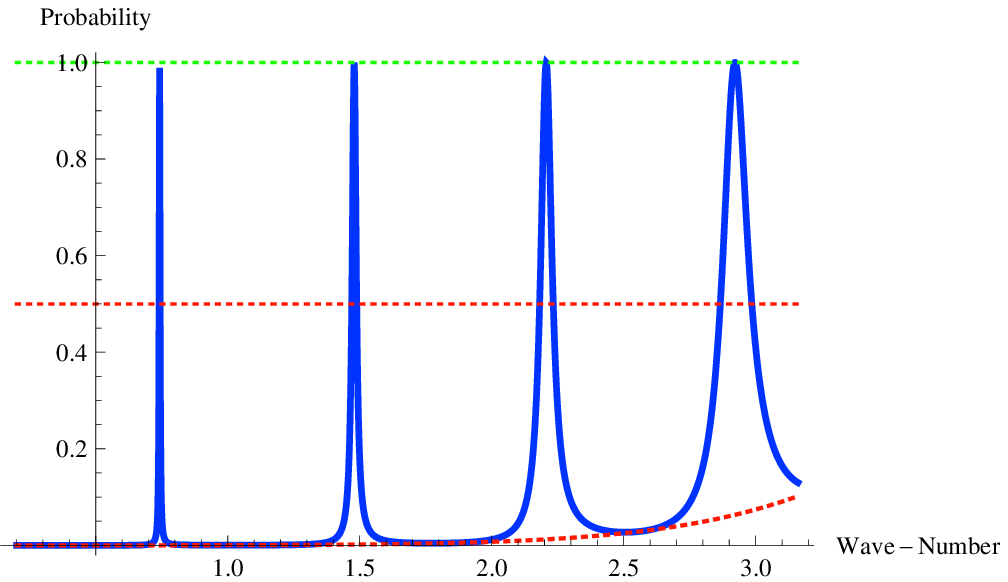}}
   \caption{Transmission rate plotted as a function of wave number $k$. (a). The relevant parameters for optical Fabry-Pe$\acute{r}$ot interferometer are $R=80\%,d_0=2\text{cm}$ and wave number $k$ is in unit $10^6/\text{m}$. (b). The relevant parameters for a symmetric double-barrier are $A_1=A_2=10.36\text{eV},~l_1=l_2=1.2{\AA}$ and $m=m_e,~d=7{\AA}$. Horizontal axis of wave number is also plotted with unit $k_e=\sqrt{2m_eE_s}/\hbar$.}\label{Compar}
\end{figure}
The two pairs of curves in the last two subgraphs of Fig.\ref{Asymm} correspond to left and right hand side of equation (\ref{localextrem}) and (\ref{resonance}) respectively, and the coincident cross points (marked with black dots) at the same wave number $k$ in these two subgraphs denote the solutions of simultaneous equations, (\ref{localextrem},\ref{resonance}). Thus these points also pin down where the resonances occur. We can utilize this pictorial method to obtain the numerical solutions of these transcendental equations, and analyze the finesse of each resonances (or filter effect) for a specific double-barrier numerically.

The finesse of an interferometer at certain wave number is defined as
\bea\label{FINE}
\mathfrak{F}_{or}=\Delta\sigma_{free}/\Delta\sigma_{FWHM},
\eea
where $\Delta\sigma_{free}$ is the distance of nearby resonances in the wave-number axis, and $\Delta\sigma_{FWHM}$ is the full width at half maximum (FWHM) of a resonance
(peak in the $k$-spectrum). For convenience, we plot transmission spectrum of optical FP interferometer and symmetric double-barrier in Fig.\ref{Compar} as a direct comparison. For a $R=80\%$ FP interferometer shown in Fig\ref{FPcurve}, the finesse defined in (\ref{FPfine}) can be easily worked out as $\mathfrak{F}=14.0496$.  While for a double-barrier matter wave interferometer, we have to work out the wave numbers at which $T_{dbs}=\hf$ is satisfied for each resonance, and this is determined by the transcendental equation
\bea\label{FWHM}
M^2[\cosh(2\beta{l})-1]\left\{1+\frac{M^2}{2}[\cosh(2\beta{l})-1]\right\}[1+\sin(2kd+\delta)]=1.
\eea
\begin{table}[ht]
\begin{center}
\begin{tabular}{c|c|c|c|c|c|c|c|c}
\hline\hline
 Wave number at resonance  &  0.742007 &  1.47909 & 2.20664  &  2.92188  \vline\vline &  0.735410 & 1.467477 & 2.19272  & 2.907172 \\
 $\Delta\sigma_{dis}$      &  0.742007 & 0.737083 & 0.72755  & 0.71524   \vline\vline &  0.735410 & 0.732066 & 0.725246 & 0.714449\\
 \hline
 $\hf$ at left side        &  0.739935 &  1.4698  &  2.18166 &  2.86621  \vline\vline &  0.731927 & 1.450947 & 2.145098 & 2.793121\\
 $\hf$ at right side       &  0.744104 &  1.48866 &  2.23328 &  2.98555  \vline\vline &  0.738966 & 1.484995 & 2.247656 & 3.077995\\
 $\Delta\sigma_{FWHM}$     &  0.004169 &  0.01886 &  0.05162 &  0.11934  \vline\vline &  0.007039 & 0.034048 & 0.102558 & 0.284874\\
 \hline
 $\mathfrak{F}_{or}$       &  175.215  &  38.7312 &  14.1509 &  6.12092  \vline\vline &  103.252  & 21.3461  & 7.08665  & 2.55128\\
 $\mathfrak{F}_{db}$       &  177.562  &  38.8839 &  14.0283 &  6.05179  \vline\vline &  104.338  & 24.684   & 9.99817  & 4.86271  \\
\hline \hline
\end{tabular}\caption{
 This table is divided into two parts. The left four columns are the data corresponding to symmetric double barriers shown in Fig.\ref{finesse}, while the right four columns are data corresponding to asymmetric double barriers shown in Fig.\ref{ASYMDBT}. The first line in this table are wave number at which resonances occur. The second and third lines are wave numbers at which half maximum is approached from left and right in $k$-axis. The fourth line are FWHM corresponding to resonances showing in the first line. The last two lines are finesse obtained with numerical method and the analytical expression (\ref{fineTHEO}) respectively. All these numbers are in unit $k_e=\sqrt{2m_eE_s}/\hbar\approx5.12289\times10^9/\mathrm{m}$.}\label{RFDSB}
\end{center}
\end{table}

\begin{figure}
 \centering\subfigure[~Phase function $\Phi(k)$]{\label{PhaRES}\includegraphics[width=60mm]{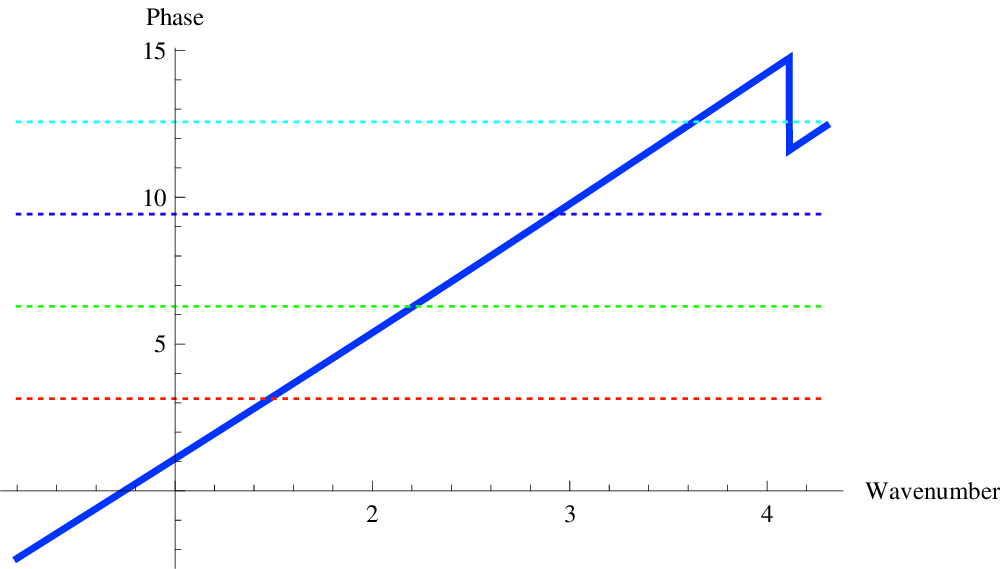}}
  \hspace{1in}
  \subfigure[~Spectrum of Asymmetric Double-Barrier]{\label{ASYMDBT}\includegraphics[width=60mm]{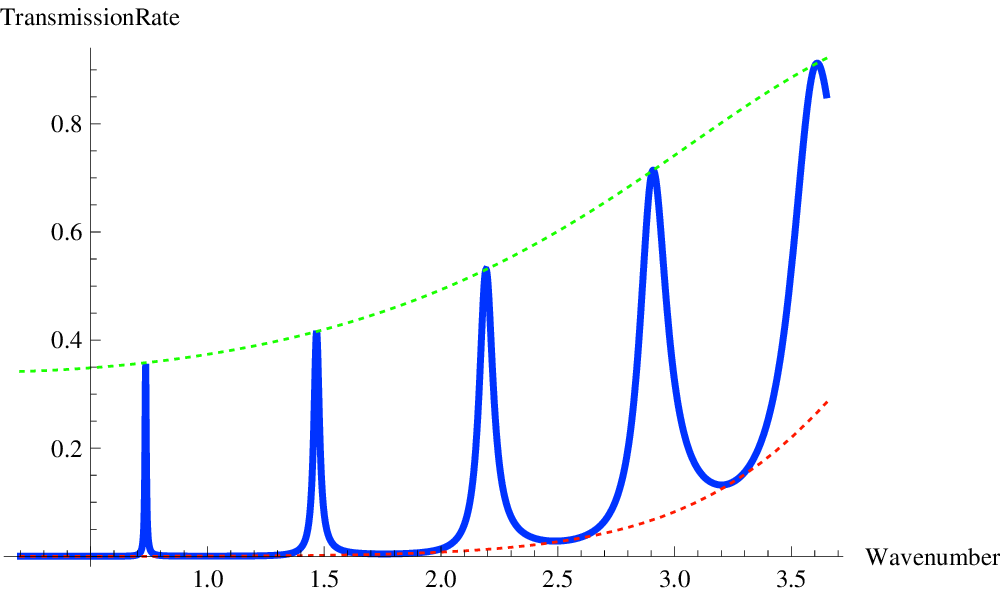}}
   \caption{(a). An approximate linear curve of $\Phi(k)$. The five dashed horizontal lines correspond to $0,\pi,2\pi,3\pi,4\pi$, while the cross points of these lines with the blue curve show out at which $k$, the resonant tunneling condition (\ref{RESC}) is satisfied. The other parameters specify the double-barrier is the same as in Fig.\ref{finesse}. (b). The transmission rate for an asymmetric barriers, with $A_1=10.6\text{eV},~l_1=1.5{\AA};A_2=8.7\text{eV},~l_1=1.{\AA}$ and $m=m_e,~d=7{\AA}$. It also show nearly four peaks in the tunneling region.}\label{MissStandPeak}
\end{figure}

For a specific double-barrier shown in Fig.\ref{finesse}, we work out the wave-numbers $k$ where the resonances occur and also the roots of transcendental equation (\ref{FWHM}) using the pictorial method shown above. The results are summarized in Table \ref{RFDSB}.
Since there are only four peaks in tunneling region, we only need to give 4-column of results. The approximate linearity of the phase $\Phi(k)\equiv{kd-\hf(\phi_1+\phi_2+\pi)}$ factor in (\ref{ResonCD}) shows that the occurrence of resonances is nearly periodic with respect to $k$ (see Fig.\ref{PhaRES}), thus $\Delta\sigma_{dis}(k)$ varies slowly and we can approximate the free distance between nearby resonances with their averages, \ie,
\bea\label{sigmafree}
\Delta\sigma_{free}=\frac{1}{4}\sum_{i=1}^4\Delta\sigma_{dis}=0.73047.
\eea
From table \ref{RFDSB} we see that $\Delta\sigma_{FWHM}$ increases with increasing wave number, hence $\mathfrak{F}_{or}$ decreases with $k$ increase. In the last two rows, we see that the finesse obtained from analytical expression (\ref{fineTHEO}), $\mathfrak{F}_{db}$, matches well with $\mathfrak{F}_{or}$ obtained numerically for the symmetric double-barrier. Even for an asymmetric double-barrier, we see the results match well for resonances in ``deep tunneling region", \ie, the fifth and sixth columns of data in table \ref{RFDSB}.

\section{Resonance and Tunneling Time}\label{3rd}
\begin{figure}
 \centering
  \subfigure[~Optical FP cavity]{\label{StandFP}\includegraphics[width=60mm]{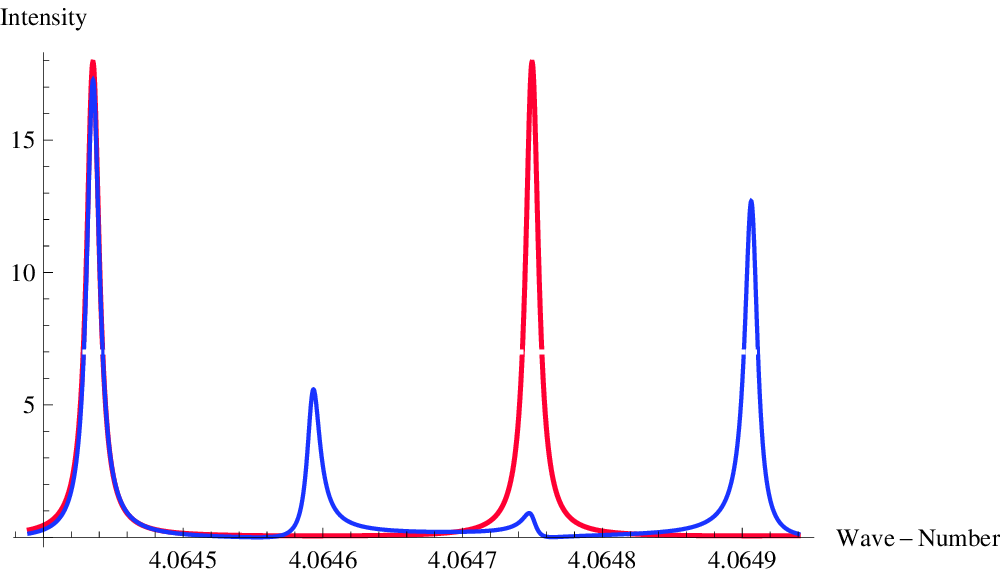}}
  \hspace{1in}
  \subfigure[~Double barrier potential]{\label{StandDB}\includegraphics[width=60mm]{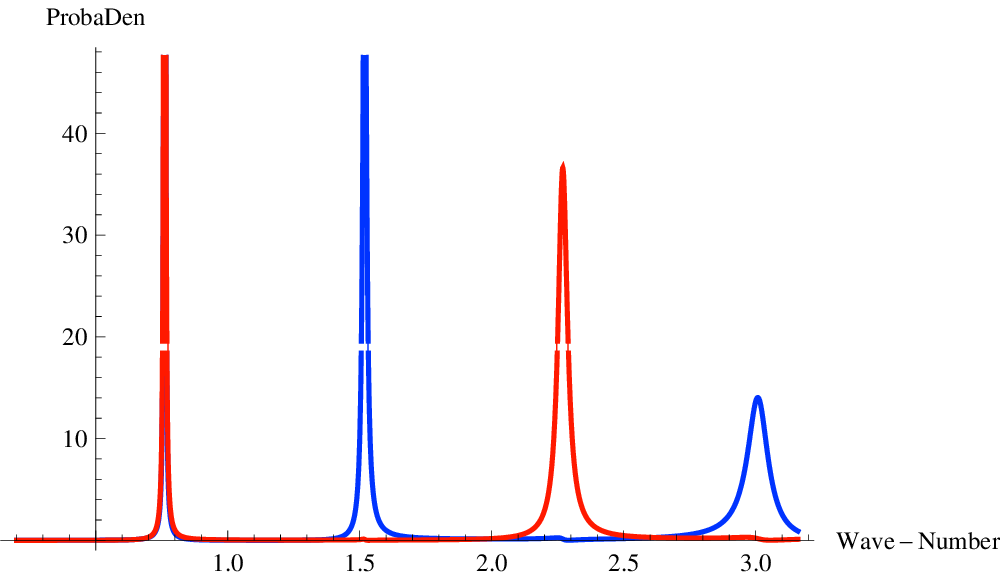}}
   \caption{$k$-spectrum of standing waves. The missing peaks are due to the presence of standing wave nodes. (a). The parameters for the spectrum of standing light wave inside a FP cavity are $R=80\%,d_0=2\text{cm}$ and $k$ is plotted with unit $10^6/\text{m}$. The blue and red curves correspond to $I(k,x)$ with position $x=5.03\text{mm}$ and $x=1\text{cm}$ respectively. (b). The parameters for the spectrum of standing matter wave of a double-barrier are $A_1=9.6\text{eV},~l_1=1.2{\AA};~A_2=25.8\text{eV},~l_2=0.8{\AA}$ and $m=m_e,~d=7{\AA}$. The blue and red curves corresponds to $P_{b_1<x<a_2}(k,x)$ with position $x=3.25{\AA}$ and $x=4.6{\AA}$ respectively. Wave number is in unit $k_e$ as mentioned above.}\label{MissStandPeak}
\end{figure}

Before investigating resonance behavior, first we consider the standing wave phenomena happening inside a FP cavity and inside a well formed between two barriers. We will see this phenomena has close relationship to resonance.
For a FP interferometer, standing wave forms due to the superposition of left moving and right moving multi-waves. The relative intensity is given by
\bea\label{SWFP}
I(k,x)=\frac{1+\mathscr{R}-2\sqrt{\mathscr{R}}\cos(2k(d_0-x))}{1-\mathscr{R}}T_{FP},
\eea
where the origin of position $x$-axis is assumed to be at the edge of left mirror of FP cavity. $I(k,x)$ as a function of $k$ is shown in Fig.\ref{StandFP}. We see that except certain missing peaks, the peaks appear nearly at the same position $k=\frac{m\pi}{d_0}$ ($m\in\mathcal{Z}$) in $k$-axis. This fact is due to that $I(k,x)$ is a product of $T_{FP}$ and a pre-factor, and the latter in general only causes a slight shift of the peaks in $k$-axis. The consequence of significant modifications caused by the pre-factor are the missing peaks. These missing peaks are due to the presence of wave nodes at certain $k$, and reflect most the standing wave nature in the cavity.
A more complete graph of $I(k,x)$ is shown in Fig.\ref{StandMisFP} in Appendix B.

Similarly, the relative probability of finding a particle in the free region ($x\in(b_1,a_2)$) between two barriers is
{\small
\bea\label{relProbStan}&&
P_{b_1<x<a_2}(k,x)=|\phi(x)|^2%=|\alpha|^2+|\gamma|^2+2Re[\alpha\gamma^*e^{2ikx}]
\nn&&~~=\{1+M_2^2[\cosh(2\beta_2l_2)-1]
+M_2[\sinh(2\beta_2l_2)\sin[2k(a_2-x)]-K_2[\cosh(2\beta_2l_2)-1]\cos[2k(a_2-x)]]\}T_{db}.
\eea
}
This time the standing matter wave is formed by the interference of left moving and right moving components, $\alpha{e}^{ikx}$ and $\gamma{e}^{-ikx}$. A 3-dimensional plot of $P_{b_1<x<a_2}(k,x)$ is shown in Fig.\ref{StandMisDB}, two slits of which are given in Fig.\ref{StandDB}. Similar to $I(k,x)$ in (\ref{SWFP}), $P_{b_1<x<a_2}(k,x)$ is also a product of transmission rate $T_{db}$ with a pre-factor, so except certain missing peaks, the relative probability approaches maximum nearly at the same $k$ where resonant tunneling happens. As is transparent in Fig.\ref{StandDB}, the pre-factor in (\ref{relProbStan}) is also responsible for the occurrence of wave-nodes (or missing peaks) and reflects the essential feature of standing wave.

Note that the wave function in (\ref{steepBarr}) is unnormalized, thus there is no need for $P_{b_1<x<a_2}(k,x)$ to be less than 1, similar to the relative intensity of a standing wave in FP cavity. Further more, due to the constructive interference effect of the matter wave components, ${\alpha}e^{ikx}$ and ${\gamma}e^{-ikx}$, $P_{b_1<x<a_2}(k,x)$ is much larger than 1 at certain resonant wave numbers.
On the other hand, there is a flux conservation condition
\bea\label{Fluxcons}
-\frac{i\hbar}{2m}[\phi(x)'\phi(x)^*-{\phi(x)^*}'\phi(x)]\equiv\text{Const},
\eea
which insures that
\bea\label{Fluxconrt}&&
|\mathcal{T}_{db}|^2=|\alpha|^2-|\gamma|^2, \quad  |\mathcal{T}_{db}|^2+|\mathcal{R}_{db}|^2=1.
\eea
So the net flux in this 1-dimensional problem is a right moving one, together with the reflected left moving flux, the total flux is always conserved.\\

Next we turn to the tunneling time of double-barrier, this may shed some new light on resonance effect. As a physical process, it must take certain time for a particle to tunnel through a barrier. However, time in quantum mechanics is a rather controversial subjects\cite{TIQM}. Up to now, there does not exist a well accepted definition on tunneling time in the literature, though there does have several well-known proposals\cite{Peres}\cite{TTCR}\cite{BT} and a tremendous amount of works on this subject\cite{Winful}\cite{ROTT}\cite{Packet}(for details, see \cite{TIQM}\cite{contin}). Here we do not try to give a self-contained discussion or clarification on this obscure subject, instead, we just try to utilize the phase time approach to estimate how much time a resonant particle spends inside a specific double-barrier.

To make life simpler, consider the symmetric double-barrier, then the transmission coefficient (\ref{TranscDSPS}) can be rewritten as
\bea\label{PhasTSC}
\mathcal{T}_{dbs}=|\mathcal{T}_{dbs}|\exp[i\delta_p(k;\mathcal{O})],
\eea
where $\delta_p\equiv\theta(k;\mathcal{O})-k(2l+d)$ and $\theta(k;\mathcal{O})$ is the phase shift for a particle tunneling through double barriers. Here $\mathcal{O}$ denotes all the other parameters (like barrier length $l$, distance between two barriers $d$) except the wave number. The phase shift can be written explicitly as
\bea\label{PhaseST}
\theta(k;\mathcal{O})=\arctan[\frac{[M^2-K^2\cosh(2\beta{l})]\sin(kd)-K\sinh(2\beta{l})\cos(kd)}{\cosh(2\beta{l})\cos(kd)+K\sinh(2\beta{l})\sin(kd)}],
\eea
from which one can obtain the phase time
\bea\label{PhaTDB}
\tau_{dbs}=\frac{\partial\theta(k;\mathcal{O})}{\partial(k)}/\frac{dk}{dE}=\frac{m}{\hbar{k}}\frac{\partial\theta(k;\mathcal{O})}{\partial{k}}.
\eea
The explicit analytical expression of $\tau_{dbs}$ is presented in Appendix C. Before analyzing $\tau_{dbs}$, we think it is valuable to review the result of single barrier first, and then compare it with that of double-barrier.
The phase shift for a single barrier is
\bea\label{PhaseSTi}
\theta_{sb}=\arctan[\frac{K\sinh(\beta{l})}{\cosh(\beta{l})}],
\eea
from which we can get the corresponding phase time
\bea\label{PhaTSB}
\tau_{sb}=\frac{m}{\hbar{k}}\frac{M^2\sinh(2\beta{l})+K(kl)}{\beta[1+M^2\sinh(\beta{l})^2]}.
\eea

\begin{figure}
 \centering\subfigure[~$\tau_{sb}$ as a function of $k$ and $L$]{\label{tauSBkL}\includegraphics[width=65mm]{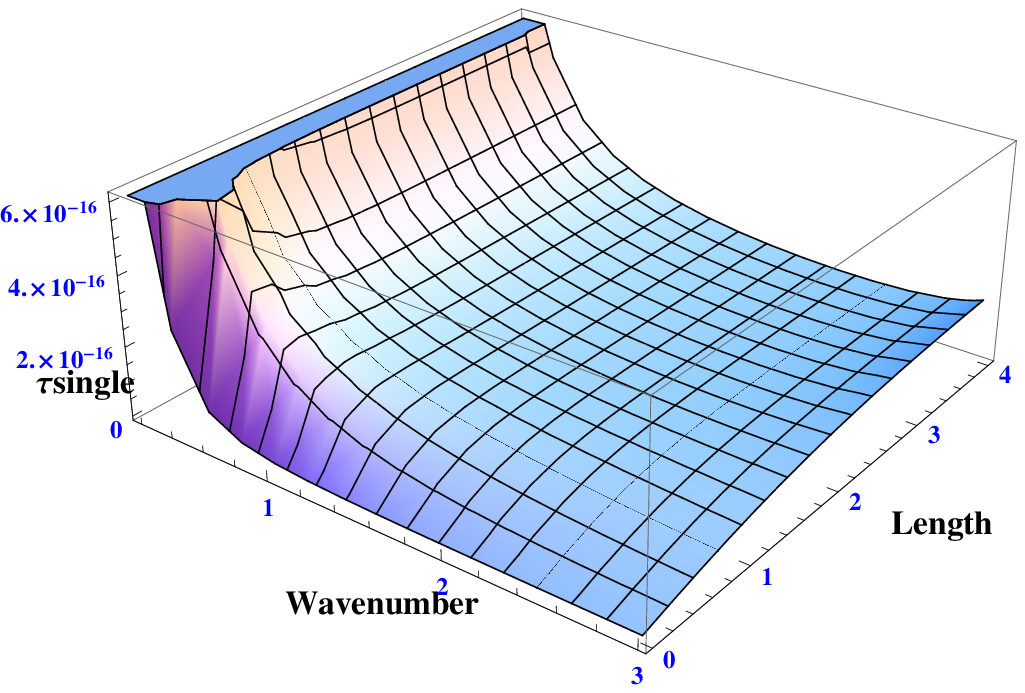}}
  \hspace{0.1in}
  \subfigure[~$\tau_{dbs}$ as a function of $k$ and $A$]{\label{tauSBkA}\includegraphics[width=80mm]{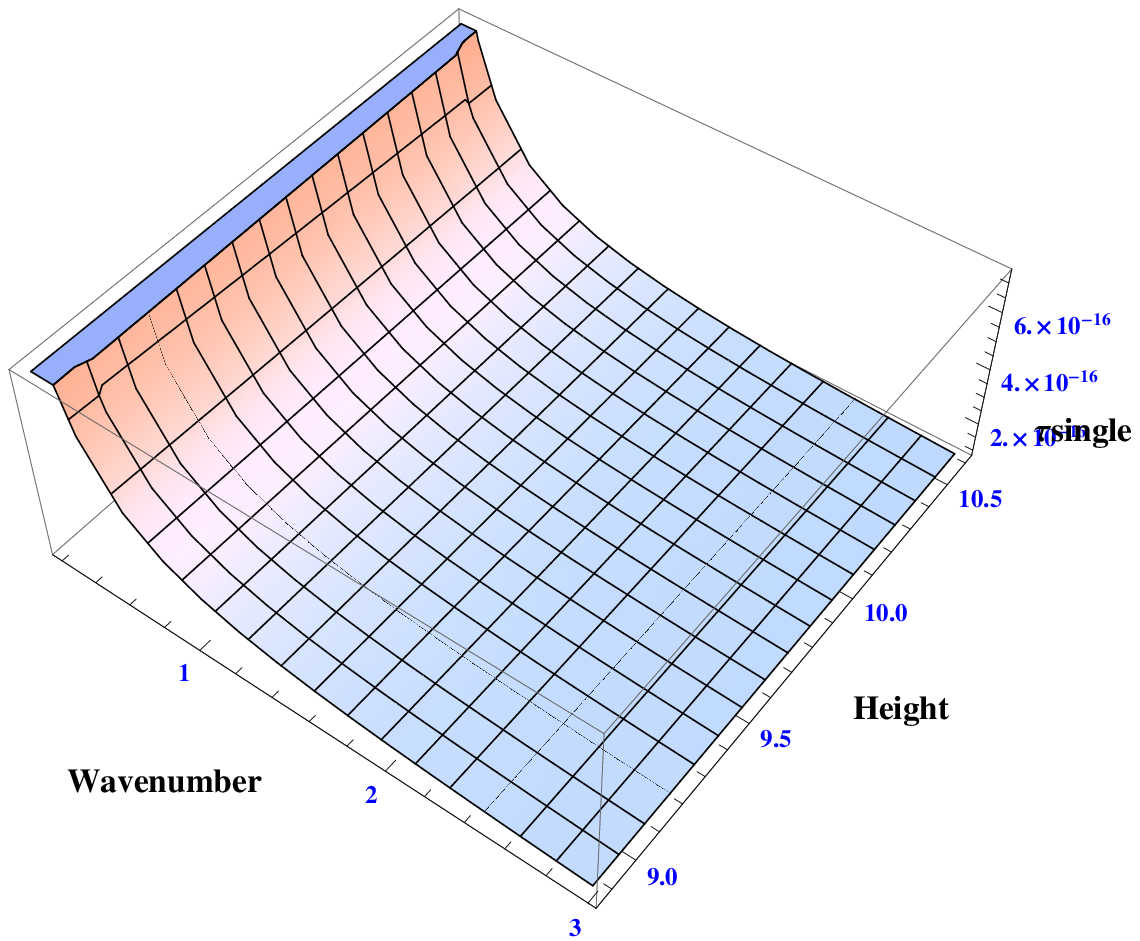}}
  \caption{Wave number is plotted with unit $k_e$ as mentioned above. (a). The 3-Dim graph is specified with barrier height $A=10.36$eV. (b). The 3- Dim graph is specified with barrier length $l=1.2{\AA}$.}\label{Timelag}
\end{figure}

We show in Fig.\ref{tauSBkL} and Fig.\ref{tauSBkA} that how $\tau_{sb}$ varies with respect to its arguments, wave number $k$, barrier length $l$ and barrier height $A$.
At much small length $l$, $\tau_{sb}$ decreases monotonously with increasing wave number. This is anticipated, since energetic particle moves faster
than less energetic one, obviously it will take less time for an energetic particle to cross over the barrier. At larger length $l$, $\tau_{sb}$
decreases with increasing $k$ for $k<k_0$, and increases slowly for $k>k_0$, where $k_0$ is a stationary wave number satisfying $\frac{\partial\tau_{sb}}{\partial{k}}|_{k=k_0}=0$.
%(for much larger $l$, $k_0$ is slightly larger than $\sqrt{2mA/3}/\hbar$).
The barrier height dependence of $\tau_{sb}$ is quite similar. At rather small $l$, to cross over a higher barrier, a particle with a certain $k$ will spend more time, while for larger $l$, phase time decreases with increasing $A$.
Generally speaking, thin barriers slow down particle's velocity, while for thick barriers, the more repulsive the barrier is (\ie, the larger barrier area $Al$), the less time it will spend in tunneling.
In other words, $\tau_{sb}$ approaches $\frac{l}{v}(2+A/E)$ when $\beta{l}\ll1$ and saturates at the value $\frac{2}{v\beta}$ when $\beta{l}\gg1$.
The saturation behavior of $\tau_{sb}$ with length $l$ can be apparently seen in Fig.\ref{tauSBkL} and is a signal of the well-known superluminal phenomena discussed in the literature\cite{superluminal,Hartman}. However, casuality isn't  violated as shown in\cite{causality}\cite{Davies}. We suspend our review here and return to the discussion of double-barrier.

\begin{figure}
 \centering\subfigure[~$\tau_{dbs}$ as a function of $k$ and $d$]{\label{tauDBkd}\includegraphics[width=60mm]{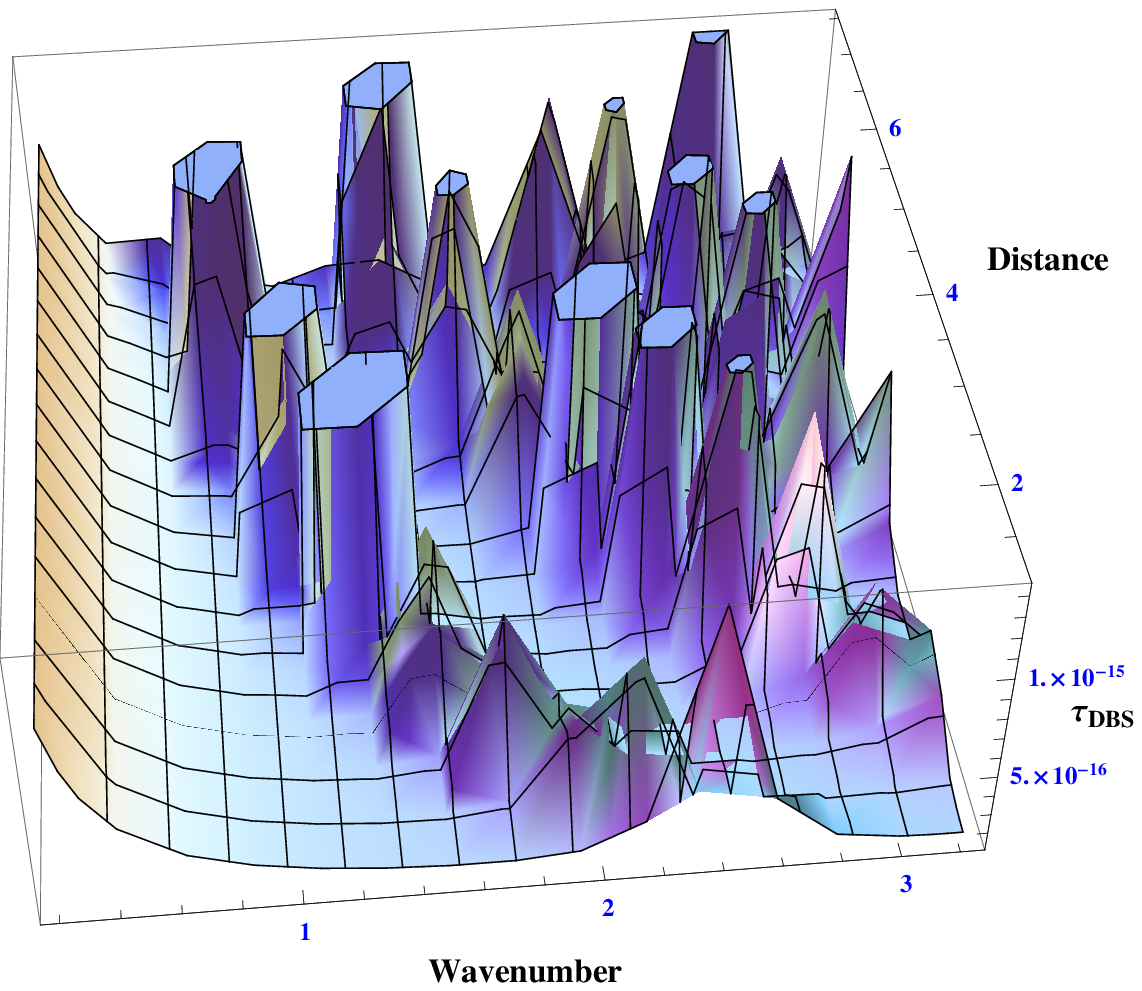}}
  \hspace{0.1in}
  \subfigure[~$\tau_{dbs}$ as a function of $k$ and $L$]{\label{tauDBkL}\includegraphics[width=70mm]{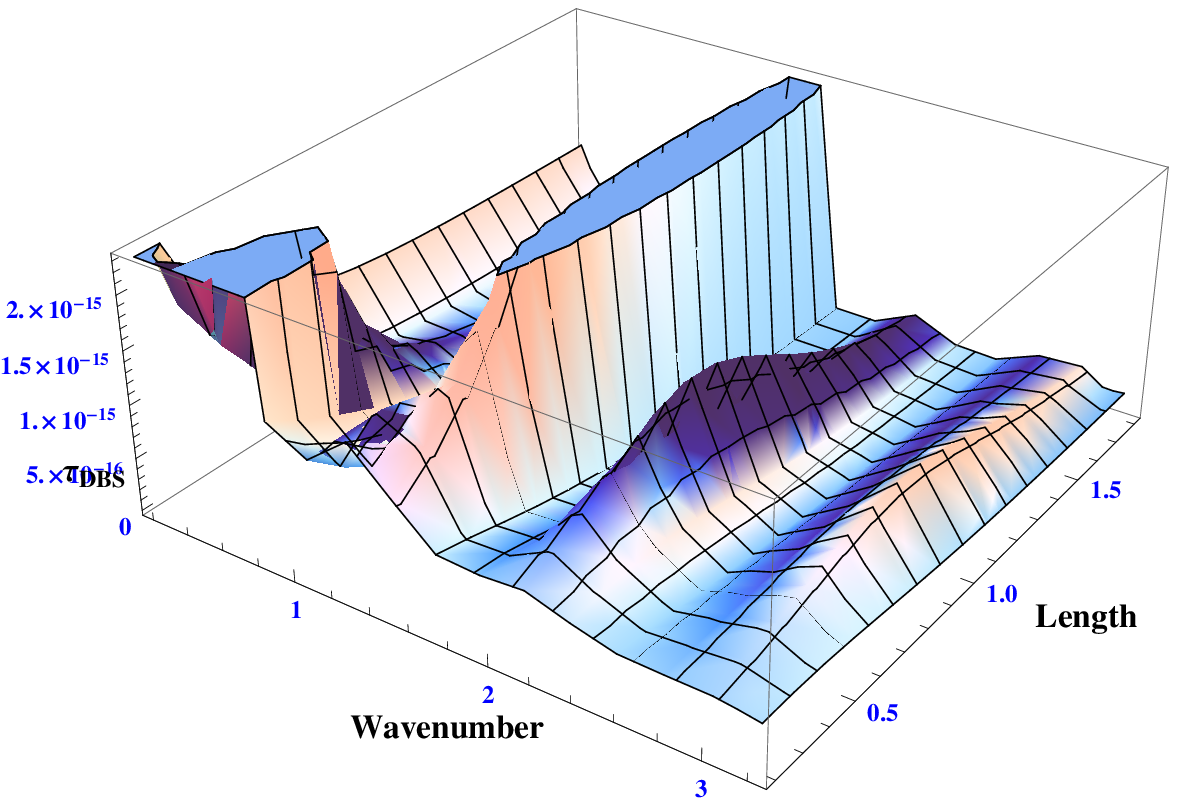}}
  \caption{Wave number is plotted with unit $k_e$ as mentioned above. (a). The 3-Dim graph is specified with barrier height $A=10.36$eV and barrier length $l=1.2{\AA}$. (b). The 3-Dim graph is specified with barrier height $A=10.36$eV and distance between two barriers $d=7{\AA}$.}\label{Timelag}
\end{figure}

% $d\ll\frac{\pi}{4k_t}$
At much small barrier distance $d$ and small barrier area (\ie, small $A$ and $l$), the phase time behavior of double-barrier with barrier length $l$ is qualitatively the same with that of a single barrier with length $2l$. This similarity is not surprising since symmetric double barriers with barrier length $l$ becomes single barrier with length $2l$ in the limit $d\rightarrow0$. Continuity guarantees this resemblance in the case of small barrier distance $d$ and barrier area $Al$. This can be confirmed from the comparison of Fig \ref{tauDBkL} with Fig \ref{tauSBkL}.

This resemblance breaks down with larger barrier distance. In this case, the presence of a well inside two barriers renders this time behavior quite different from that of a single barrier. From the analytic expression
\bea\label{DphaT}
\tau_{dbs}=\frac{m}{\hbar}\frac{\mathcal{D}[k,\beta,l,d]}{k}T_{dbs},
\eea
we see the effective length of tunneling region is a product of tunneling rate $T_{dbs}$ and $\mathcal{D}[k,\beta,l,d]$ ($\mathcal{D}[k,\beta,l,d]$ is defined in Appendix C), both of which contain periodic sine functions of $kd$, thus it is obvious that $\tau_{dbs}$ must have certain resonant behaviors. This is visualized in Fig.\ref{tauDBkd} with Fig\ref{tauDBkL}. The qualitative behavior can be summarized as $\tau_{dbs}\approx\{d+2l+2lM[\frac{\beta}{k}+M(kl)\sin(2kd)]\}/v$ when $\beta{l}\ll1$ and $\tau_{dbs}\approx\frac{2/\beta}{v}$ when $\beta{l}\gg1$ (where $v=\hbar{k}/m$). Interestingly, the phase time of both double-barrier and single barrier saturate at the same limit when $\beta{l}\gg1$, so the superluminal problem haunts us again in the double-barrier case. This is not surprising, since in the limit $l\rightarrow\infty$, the presence of a finite well is no longer important, $\tau_{dbs}$ thus approaches the same limit as that of $\tau_{sb}$. We will not hesitate here about the subtle causality problem in quantum tunneling. Rather, we turn to the more practical and interesting analysis of resonant behaviors of $\tau_{dbs}$ instead.

As is apparent in Fig.\ref{tauDBkd}, the presence of a well is responsible for the landscape of peaks and valleys in phase time of double-barrier. Generally speaking, for a given double-barrier heights, a wider well allows the formation of more peaks, hence resonance. This fact is not only reflected in tunneling rate $T_{db}$ (or $T_{dbs}$) and $P_{b_1<x<a_2}(k,x)$, but also in phase time (\ref{DphaT}). So we guess the sharp peaks in tunneling time can be attributed to the longer life-time of resonances.
As a numerical check, we use uncertainty principle
\bea\label{UCP}
\delta{t}\sim\hbar/\delta{E}
\eea
to give an estimate of the life-time of resonances for a specific double-barrier.
since we have already calculated $\Delta\sigma_{FWHM}$ in table \ref{RFDSB}, it is easy to obtain the width of resonance by
$\delta{E}\sim\hbar^2k\delta{k}/m$,
where we have approximated $\delta{k}$ with $\Delta\sigma_{FWHM}$.
\begin{table}[ht]
\begin{center}
\begin{tabular}{c|c|c|c|c|c|c|c|c}
\hline\hline
 Wave number at resonance  &  0.742007  &  1.47909  &  2.20664  &  2.92188 \vline\vline &  0.735410 & 1.467477 & 2.19272  & 2.907172 \\
 \hline
 $\Delta\sigma_{FWHM}$     &  0.004169  &  0.01886  &  0.05162  &  0.11934 \vline\vline &  0.007039 & 0.034048 & 0.102558 & 0.284874\\
 \hline
 $2\tau_{uc}$              &  213.041   &  23.6247  &  5.78567  &  1.88997 \vline\vline &  127.31   & 13.1899  & 2.93055  & $7.95755*10^{-1}$\\
 $\tau_{db}$              &  213.311   &  23.8112  &  5.95756  &  2.08377 \vline\vline &  134.175  & 15.1924  & 4.33754  & 1.73741 \\
\hline \hline
\end{tabular}\caption{$\tau_{uc}$ is the life time calculated from uncertainty principle, while $\tau_{db}$ is that obtained directly from phase time. As in the table \ref{RFDSB}, the first four column are time quantities for resonance of symmetric double barriers while the last four column are that of asymmetric double barriers. The time unit is in femtosecond ($10^{-15}$s).}\label{lifetime}
\end{center}
\end{table}
The results together with that calculated from (\ref{DphaT}) is presented in Table \ref{lifetime}.

We see that these two estimates match very well with an error less than $0.3$fs for symmetric double barriers. Using semi-analytical method, we also obtain the phase time of an asymmetric double-barrier with parameters specified in Fig.\ref{ASYMDBT}. We see that for high finesse resonance, the life time estimated by phase time approach and that with uncertainty principle coincide well at least in the order of magnitude. Of course, if the estimation of life-time is replaced by $\tau_{db}(k,d,l_1,l_2)-[\tau_{sb}(k,l_1)+\tau_{sb}(k,l_2)]$, the results will be more close to $2\tau_{uc}$. The factor of 2 in $2\tau_{uc}$ is inserted to fit the data well.
Up to now, at least in the order of magnitude estimation, we do provide an alternative method, though we have not worked out why the phase time approach can be a good evaluation of the lifetimes of resonances in double-barrier. Probably the fact that the energy eigenvalue of bound states in a well with width $d$ formed by two infinite thick barriers with height $A$ satisfy $M^2\sin^2[kd]=1$ may provide a hint of the underline reason. The detail analysis of this problem is beyond the scope of this paper.

\section{Conclusion}\label{4th}
We have revisited the problem of quantum tunneling through a double-barrier with a close analogy of optical Fabry--P$\acute{e}$rot (FP) interferometer. Though there have many interesting schemes of matter wave FP interferometer\cite{ChamonWen}\cite{EHallFP}\cite{CQFP}\cite{TDAFP}, and the physics involved are much richer than double barriers, our derivation of transmission and reflection coefficients are based on a common concept between the two analogs, multi-wave interference. Not only our derivation relies only on a knowledge of reflection and transmission coefficients through two single barriers (without solving the Schr$\ddot{o}$dinger equation of double barriers), but utilizing the same analogy, we also obtain a  finesse formula (\ref{fineTHEO}), which is in good accordance with the definition, (\ref{FINE}), for resonances in ``deep tunneling" region, especially for the symmetric double barriers.

Then we discuss the relevant standing wave phenomena, which is also a common feature of FP interferometer and double-barrier tunneling. Moreover, it is associated to the occurrence of resonance. Utilizing phase time approach, we calculate the phase time of double barriers in comparison with that of a single barrier. The superluminal phenomena or Hartman effect\cite{Hartman} also resides in the double barrier case, which may be a general feature of quantum tunneling. Of course, several suggestions have been proposed to mitigate or even eliminate this effect and the associated causality problems\cite{causality}\cite{Davies}\cite{SWP2011}. Here we don't discuss this topic in this paper, we feel that a suitable solution should be applicable to various barrier shapes and may even involve relativistic quantum formalism. Tunneling time of various shapes of double barriers has already been discussed in the literature\cite{PPODB}\cite{PaODB}\cite{SWP2011} and this paper only provides a slight touch on phase time of double square barriers. We find phase time of both symmetric double-barrier and single barrier saturate at the same limit $\frac{2/\beta}{v}$ when $\beta{l}\rightarrow\infty$. Further we focus our attention on the peaks and valleys in the configuration of double-barrier phase time. With a numerical support, we interpret these peaks as results of longer life time of resonances formed inside the well between two barriers. The numerical results match well with that estimated from uncertainty principle, especially for resonance with high finesse. Thus we think phase time may provide a helpful method to the evaluation of resonance's life time. Furthermore, we even guess that other time quantities, like dwell time, Larmor time, \etc\cite{TTCR}\cite{BT}, may have interesting applications in the estimate of life-time of resonances in quantum tunneling. Though the underline relation of phase time and the life time of resonance is not clear, we think it is valuable and interesting to explore it in the near future.

\section*{Acknowledgments}
This work was supported by the National Natural Science Foundation of China (Grant No. 61203187).
Zhi-X wishes to thank Jing-ming Song for useful discussions.

\section{Appendix A}\label{App1st}
Various parameters characterizing tunneling through double potential barriers
\bea\label{DPB2}&&
U_{db}(x)=A_1[\Theta(x-a_1)\Theta(b_1-x)]+A_2[\Theta(x-a_2)\Theta(b_2-x)],\quad~\text{with}~~b_i-a_i=l_i,~i=1,2;\quad a_2-b_1=d,
\eea
with wave number $k\equiv\frac{\sqrt{2mE}}{\hbar}$ and $\beta_i\equiv\frac{\sqrt{2m(A_i-E)}}{\hbar}$ ($i=1,2$) is given in this section. To make the expressions as compact as possible, we define
\bea\label{defMK}
M_i\equiv\hf(\frac{\beta_i}{k}+\frac{k}{\beta_i}), \quad K_i\equiv\hf(\frac{\beta_i}{k}-\frac{k}{\beta_i});  ~~~(i=1,2).
\eea
Then the transmission coefficient $\mathcal{T}_{db}$ in (\ref{Tdcoeff}) is
\bea\label{Tcoeffd}&&
\mathcal{T}_{db}=\frac{e^{-ik(d+l_1+l_2)}}{\mathcal{F}[k;\beta_1,\beta_2;d;l_1,l_2]},\\&&
\mathcal{F}[k;\beta_1,\beta_2;d;l_1,l_2]=e^{ikd}[M_1\sinh(\beta_1{l_1})][M_2\sinh(\beta_2{l_2})]+
e^{-ikd}[\cosh(\beta_1{l_1})+iK_1\sinh(\beta_1{l_1})]\nn&&
~~~~~~~~~~~~[\cosh(\beta_2{l_2})+iK_2\sinh(\beta_2{l_2})].
\eea
The other coefficients in (\ref{ACoeff}) are
\bea\label{tunnel2d}&&
c_2=\frac{1}{2}e^{-\beta_2(d+l_1+l_2)}(1+\frac{ik}{\beta_2})\mathcal{F}^{-1}e^{(ik-\beta_2)a_1}, \quad  d_2=\frac{1}{2}e^{\beta_2(d+l_1+l_2)}(1-\frac{ik}{\beta_2})\mathcal{F}^{-1}e^{(ik+\beta_2)a_1};\\&&\label{transd}
\alpha=e^{-ik(d+l_1)}[\cosh(\beta_2{l_2})+iK_2\sinh(\beta_2{l_2})]\mathcal{F}^{-1}, \quad
\gamma=-ie^{ik(d+l_1+2a_1)}M_2\sinh(\beta_2{l_2})\mathcal{F}^{-1};\\&&\label{tunnel1d}
c_1=\frac{1}{2}e^{-\beta_1l_1}\{(1+\frac{ik}{\beta_1})e^{-ikd}[\cosh(\beta_2{l_2})+iK_2\sinh(\beta_2{l_2})]-i(1-\frac{ik}{\beta_1})e^{+ikd}M_2\sinh(\beta_2{l_2})\}\mathcal{F}^{-1}e^{(ik-\beta_1)a_1},\\&&
d_1=\frac{1}{2}e^{\beta_1l_1}\{(1-\frac{ik}{\beta_1})e^{-ikd}[\cosh(\beta_2{l_2})+iK_2\sinh(\beta_2{l_2})]-i(1+\frac{ik}{\beta_1})e^{+ikd}M_2\sinh(\beta_2{l_2})\}\mathcal{F}^{-1}e^{(ik+\beta_1)a_1};\\&&
\label{Rcoeffd}
\mathcal{R}_{db}=e^{-i\frac{\pi}{2}}\mathcal{F}^{-1}\{e^{-ikd}[\cosh(\beta_2{l_2})+iK_2\sinh(\beta_2{l_2})]M_1\sinh(\beta_1{l_1})+e^{ikd}[\cosh(\beta_1{l_1})-iK_1\sinh(\beta_1{l_1})]M_2\sinh(\beta_2{l_2})\}e^{2ika_1},\nn
\eea
where $\mathcal{F}=\mathcal{F}[k;\beta_1,\beta_2;d;l_1,l_2]$.
\section{Appendix B}\label{App2nd}
In this section, we give a detail derivation of reflective coefficient from the multi-wave interference picture. The reflective coefficient as the sum of wave components is
\bea\label{Rdcoef}&&
\mathcal{R}_{db}=\sum_{i=1}^{+\infty}r_i=R_1\exp[i\phi_{r1}]+T_1^2R_2\exp[i(2\phi_{t1}+\phi_{r2})+i2k(d+l_1)]+T_1^2R_2^2R_1\exp\{i[2(\phi_{t1}+\phi_{r2})+\phi_{r1}]+i2k(2d+l_1)\}\nn&&
~~~~~~+T_1^2R_2(R_2R_1)^2\exp\{i[(2\phi_{t1}+\phi_{r2})+2(\phi_{r1}+\phi_{t2})]+i2k(3d+l_1)\}+\ldots\nn&&
~~~~=R_1\exp[i\phi_{r1}]+\frac{T_1^2R_2\exp[i(2\phi_{t1}+\phi_{r2})+i2k(d+l_1)]}{1-R_1R_2\exp[i(\phi_{r1}+\phi_{r2}+2kd)]}\nn&&
~~~~=R_1\exp[i\phi_{r1}]-\frac{T_1\frac{R_2}{T_2}\exp\{i[2(\phi_{t1}+kl_1)-\phi_{r1}+kd]\}}{\frac{R_1R_2}{T_1T_2}\exp(ikd)-\frac{1}{T_1T_2}\exp[-i(\phi_{r1}+\phi_{r2}+kd)]}\nn&&
~~~~=R_1\exp[i\phi_{r1}]-\frac{(R_2/T_2)T_1e^{i\phi_{t1}}\exp\{i[\phi_{t1}-\phi_{r1}+k(2l_1+d)]\}}{\mathcal{F}[k;\beta_1,\beta_2;d;l_1,l_2]},
\eea
where in the last step, we used the identities led to eqn. (\ref{redTdco}). Then we can quickly identify
\bea\label{idenT}&&
T_1e^{i(\phi_{t1}+kl_1)}=\frac{\cosh(\beta_1{l_1})-iK_1\sinh(\beta_1{l_1})}{1+M_1^2\sinh^2(\beta_1{l_1})},\quad \phi_{t1}-\phi_{r1}+kl_1=\frac{\pi}{2}\\&&\label{idenR}
R_1\exp[i\phi_{r1}]=\frac{-iM_1\sinh(\beta_1{l_1})}{\cosh(\beta_1{l_1})+iK_1\sinh(\beta_1{l_1})},\quad R_2/T_2=M_2\sinh(\beta_2{l_2})
\eea
by utilizing (\ref{Tsrate},\ref{Rsrate}) and (\ref{Tspha},\ref{Rspha}).
Substituting these equations (\ref{idenT},\ref{idenR}) back into (\ref{Rdcoef}), we can finally get
\bea\label{Rdcoeff}&&
\mathcal{R}_{db}=-i\frac{\frac{M_1\sinh(\beta_1{l_1})\mathcal{F}[k;\beta_1,\beta_2;d;l_1,l_2]}{\cosh(\beta_1{l_1})+iK_1\sinh(\beta_1{l_1})}+\frac{M_2\sinh(\beta_2{l_2})}
{\cosh(\beta_1{l_1})+iK_1\sinh(\beta_1{l_1})}e^{ikd}}{\mathcal{F}[k;\beta_1,\beta_2;d;l_1,l_2]}\nn&&
~~~~=-i\frac{M_1\sinh(\beta_1{l_1})\mathcal{F}[k;\beta_1,\beta_2;d;l_1,l_2]+M_2\sinh(\beta_2{l_2})e^{ikd}}{[\cosh(\beta_1{l_1})+iK_1\sinh(\beta_1{l_1})]\mathcal{F}[k;\beta_1,\beta_2;d;l_1,l_2]}\nn&&
~~~~=-i\frac{M_1\sinh(\beta_1{l_1})[\cosh(\beta_2{l_2})+iK_2\sinh(\beta_2{l_2})]e^{-ikd}+M_2\sinh(\beta_2{l_2})[\cosh(\beta_1{l_1})-iK_1\sinh(\beta_1{l_1})]e^{ikd}}{\mathcal{F}[k;\beta_1,\beta_2;d;l_1,l_2]},
\eea
which coincides exactly with (\ref{Rcoeffd}).

\section{Appendix C}\label{App3rd}
In this appendix, we present the analytical expression of phase time of square double-barrier potential. The phase shift is defined as $\theta_{dbs}$ with
\bea\label{PhaseST}
\tan[\theta_{dbs}]=\frac{[M^2-K^2\cosh(2\beta{l})]\sin(kd)-K\sinh(2\beta{l})\cos(kd)}{\cosh(2\beta{l})\cos(kd)+K\sinh(2\beta{l})\sin(kd)}.
\eea
So the derivative of $\theta_{dbs}$ with respect to wave number $k$ is
\bea\label{dePhaWNk}
\frac{\partial\theta_{dbs}}{\partial{k}}=\frac{\partial\tan[\theta_{dbs}]}{\partial{k}}/[1+\tan[\theta_{dbs}]^2]=\frac{\mathcal{D}[k,\beta,l,d]}{\mathcal{S}[k,\beta,l,d]},
\eea
where $\mathcal{S}[k,\beta,l,d]=|\mathcal{F}[k,\beta,l,d]|^2$ in (\ref{TdrateS2^-1}) and
\bea\label{Ddepha}&&
\mathcal{D}[k,\beta,l,d]=d[M^2\cosh(2\beta{l})-K^2]+\frac{2}{\beta}[M^2\sinh(2\beta{l})+K(kl)][1+\frac{M^2}{2}(\cosh(2\beta{l})-1)]+\nn&&
~~\frac{2M^2}{\beta}\left\{(\cosh(2\beta{l})-1)[(1-\frac{M^2}{2})\sinh(2\beta{l})-\frac{Kkl}{2}]\cos(2kd)+[K\cosh(2\beta{l})(\cosh(2\beta{l})-1)+\frac{\sinh(2\beta{l})kl}{2}]\sin(2kd)\right\}.\nn
\eea
Thus the analytical expression of tunneling time in the stationary phase approximation is

%\bea\label{dtan}&&
%[\cosh(2\beta{l})\cos(kd)+K\sinh(2\beta{l})\sin(kd)]^2\frac{\partial\tan[\delta_{dbs}]}{\partial{k}}=d[M^2\cosh(2\beta{l})-K^2]\nn&&
%~~~~+\frac{2}{\beta}\{\sin[kd]^2\left[M^2K^2\sinh(2\beta{l})(\cosh(2\beta{l})-2)
%+[1+M^2(\cosh(2\beta{l})-1)]K(kl)+M^4\sinh(2\beta{l})\right]\nn&&
%~~~~+\cos[kd]^2[M^2\sinh(2\beta{l})\cosh(2\beta{l})+K(kl)]+\frac{\sin[2kd]}{2}[2M^2K\cosh(2\beta{l})(\cosh(2\beta{l})-1)+M^2\sinh(2\beta{l})(kl)]\}.\nn
%\eea
%Substituting (\ref{dtan}) and (\ref{PhaseST}) into (\ref{ddeltak}), one can easily get an analytical expression of tunneling induced phase time
\bea\label{PPhaT}
\tau_{dbs}=\frac{m}{\hbar}\frac{\mathcal{D}[k,\beta,l,d]}{k}T_{dbs}.
\eea

\begin{figure}
 \centering\subfigure[~Standing Wave in FP Cavity]{\label{StandMisFP}\includegraphics[width=70mm]{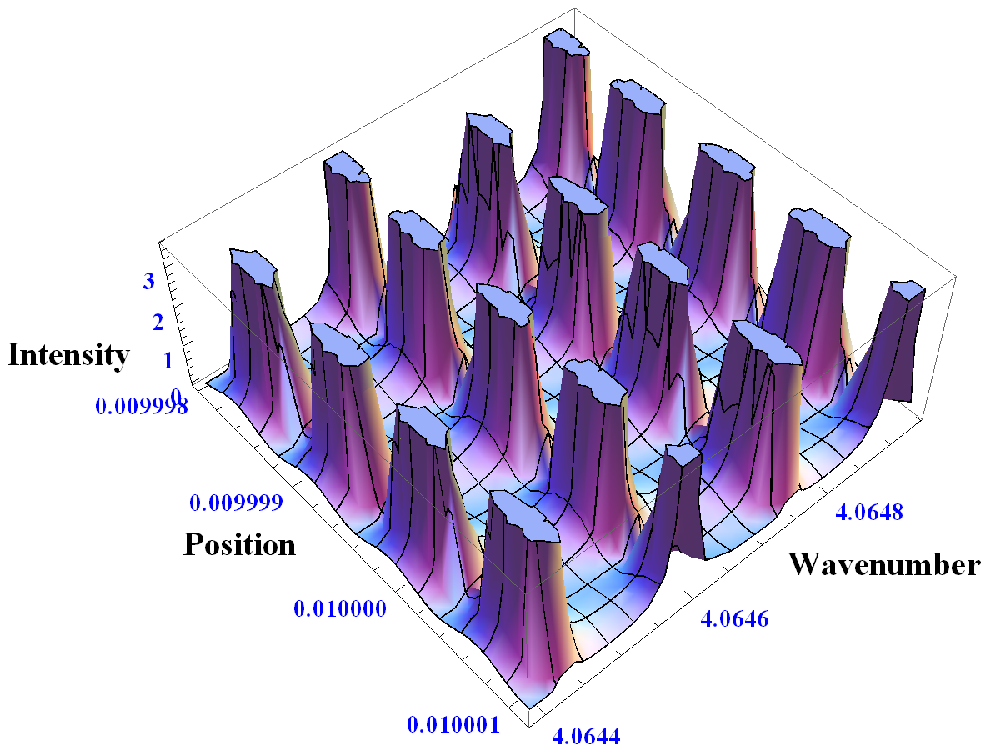}}
  \hspace{0.1in}
  \subfigure[~Standing Wave in Vally between Double Barriers]{\label{StandMisDB}\includegraphics[width=65mm]{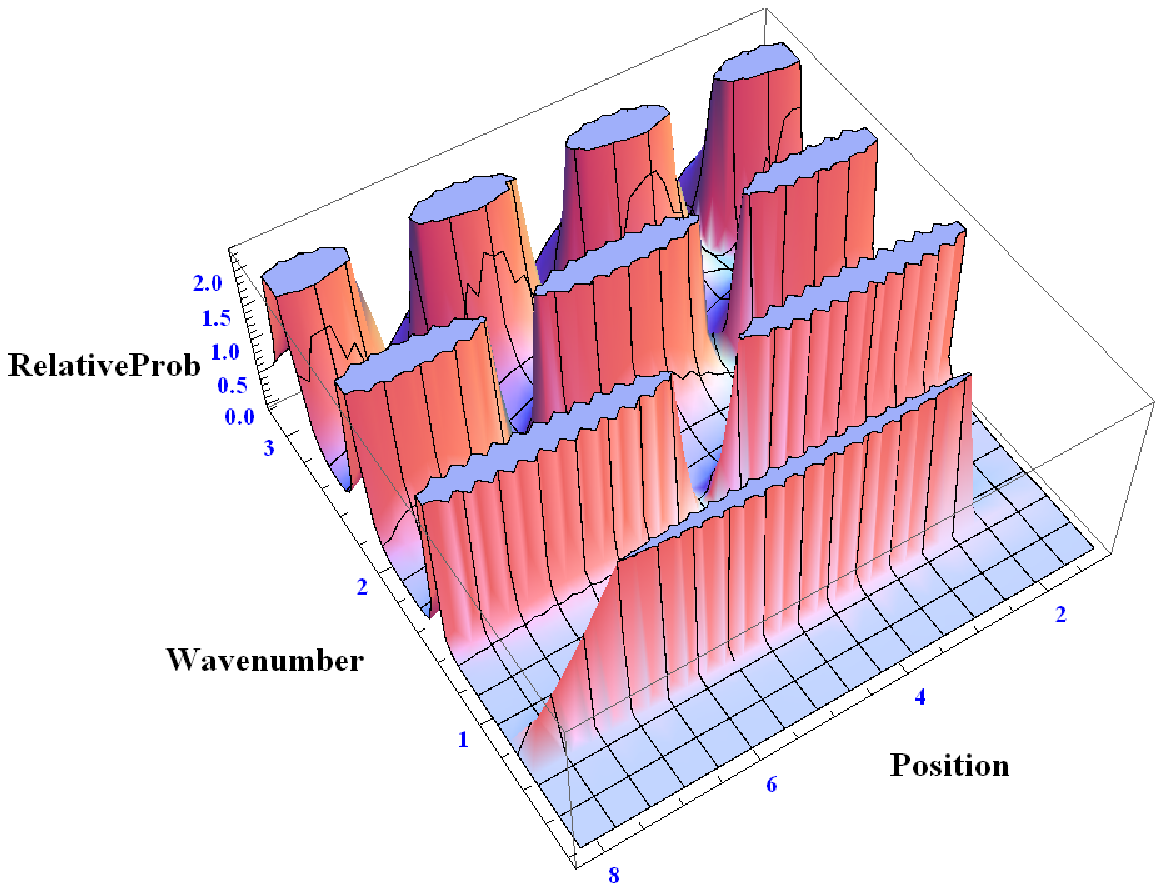}}
  \caption{(a). The relative intensity $I(k,x)$ of a standing wave in FP cavity, with $R=0.80$ and $d=2$cm. Wave number $k$ is plotted with unit $10^6/$m. (b). The relative probability $P(k,x)$ of a standing matter wave in the well of a symmetric double barrier, with parameter specified as $A=10.36$eV, $l=1.2{\AA}$ and $d=7{\AA}$. Wave number is plotted with unit $k_e$ as mentioned.}\label{MissStandPeak}
\end{figure}

\end{document}